\documentclass{aa}

\usepackage{natbib,twoopt}
\usepackage[breaklinks=true]{hyperref} 
\bibpunct{(}{)}{;}{a}{}{,}             

\usepackage[varg]{txfonts}
\usepackage{graphicx}

\hypersetup{
    colorlinks,
    linkcolor=magenta,
    citecolor=blue,
    urlcolor=blue}

\begin{document}

\title{Oscillations of red giant stars with magnetic damping in the core}
\subtitle{I. Dissipation of mode energy in dipole-like magnetic fields}

\author{Jonas M\"uller\inst{1,2}
\and Quentin Copp\'ee\inst{1,2}
\and Saskia Hekker\inst{1,2}}

\institute{Heidelberger Institut für Theoretische Studien, Schloss-Wolfsbrunnenweg 35, 69118 Heidelberg, Germany
\and Zentrum für Astronomie (ZAH/LSW), Heidelberg University, Königstuhl 12, 69117 Heidelberg, Germany}

\titlerunning{Oscillations of red giant stars with magnetic damping in the core: I.}
\authorrunning{J. M\"uller et al.}

\date{Received <date> /Accepted <date>}

\abstract{
Strong magnetic fields in the core of red-giant branch stars are expected to suppress the amplitudes of the multipole modes. This occurs when the strength of the internal magnetic field approaches the critical field strength, at which the magnetic forces become comparable to the buoyancy. 
We performed Hamiltonian ray tracing simulations of magneto-gravity waves to investigate the suppression of the multipole modes in the presence of an internal dipole-like magnetic field. We took into account different stellar masses, metallicities, and ages, as well as various oscillation frequencies and spherical degrees. In particular, we estimated the trapped fraction, a measure of multipole mode suppression, which quantifies the fraction of mode energy in the core that is dissipated by the interaction with the magnetic field.
Our results indicate that the trapped fraction can be described by a simple expression, which smoothly connects the regime without multipole mode suppression with the regime with complete suppression of the multipole modes. Crucially, the trapped fraction depends only on the ratio between the strength of the internal magnetic field and the critical field strength. Therefore, our expression for the trapped fraction provides a flexible tool that can be used, for example, to estimate the amount of multipole mode suppression as a star ascends the red-giant branch or to investigate the onset of the suppression in observed power spectral densities.
}


\maketitle

\section{Introduction} \label{sect: Introduction}

In recent years, high-quality data from space-based missions such as \textit{Kepler} \citep{borucki+10} have led to major advances in the research field of asteroseismology.
Of particular interest are the oscillations of red giants (i.e., evolved low-mass stars), as their multipolar modes are sensitive to both the stellar envelope and the core. 
This is due to the mixed character of their oscillation modes, which means that they behave like a gravity mode in the stellar core and like a pressure mode in the outer layers \citep[for a review, see][]{hekker+17}. Analyzing the mixed modes can provide information about the interior of red giants, for example about the nuclear burning reaction and the evolutionary stage \citep[][]{bedding+11, mosser+11}, the core rotation rate \citep[e.g.,][]{beck+2012, mosser+12b, deheuvels+12, deheuvels+14, beck+14, gehan+18, li+24}, as well as the strength and topology of the internal magnetic field \citep{li+22, deheuvels+23, li+23, hatt+24}.

Among the more than 16000 red giants identified by \textit{Kepler} \citep[][]{yu+18}, \citet{mosser+12} have identified a population of stars that have dipole modes with unusually low amplitudes, while their radial modes appear to be normal. Since the mode amplitude is a measure for the mode energy, this indicates that the stochastic mode excitation by turbulent convection works as usual in this particular group of red giants, and suggests an additional source of damping confined to the inner regions where the radial modes are least sensitive \citep[e.g.,][]{garcia+14, fuller+15, coppee+24}. 
The dipole modes are therefore suppressed.
The red giants with suppressed dipole modes are often referred to as the "suppressed dipole mode stars".

\citet{fuller+15} proposed that the additional damping of the suppressed dipole mode stars is caused by a strong core magnetic field. Such a field could trap part of the total energy of the dipole modes within the core, thus removing it from the observable oscillations (the "magnetic greenhouse effect"). 
In the theoretical framework of \citet{fuller+15}, a dipole mode is suppressed if the frequency of the mode is lower than a characteristic frequency that depends on the strength of the internal magnetic field. For a dipole mode with a given frequency, this results in a step-like onset of the suppression as a function of the magnetic field strength.
The strength of the internal magnetic field is thus decisive for whether a mode is suppressed or not.
Red giants with normal dipole mode amplitudes are expected to have no or a weak internal magnetic field, while the suppressed dipole mode stars are expected to contain strong magnetic fields. 
Modes with a higher spherical degree can also be suppressed, although the onset of the suppression is happening at a different field strength and the effect of the suppression is less pronounced \citep{fuller+15, stello+16a, stello+16b, cantiello+16}. 

A key assumption of the mode suppression as described by \citet{fuller+15} is that all multipole mode energy that leaks into the gravity mode cavity is lost. Crucially, this means that suppressed modes cannot be mixed, as this requires a gravity mode-like behavior in the inner regions of the star. In their prescription, the suppressed modes are pure pressure modes and their amplitudes are determined by the amount of energy that does not leak into the core \citep{fuller+15}. 

This scenario is challenged by \citet{mosser+17}, who show that a significant fraction of suppressed dipole mode stars have dipole modes with a mixed character. This implies that the damping mechanism does not fully suppress the energy leaking into the gravity mode cavity. Instead, a portion of the energy must be able to leave the gravity mode cavity and contribute to the observable oscillations. Therefore, the transition between no suppression and full suppression \citep[i.e., the suppression as described by][]{fuller+15} cannot be a step function. A more gradual transition is needed to connect these two regimes.
In fact, the suppressed dipole mode stars exhibit a variety of power spectral density (PSD) morphologies, of which the one showing dipole modes of mixed character is one (Coppée et al. in prep.).

The true nature of the suppression mechanism is still under investigation. 
Magnetism is often treated as a small perturbation to the restoring forces of the oscillations \citep[e.g.,][]{prat+19,prat+20,mathis+21,bugnet+21,loi+21,li+22,bugnet22,bhattacharya+24,das+24}. However, this method can only be used for weak magnetic fields and is not applicable if the magnetic field is strong enough to cause mode suppression \citep[][]{rui+23}.
Another common assumption is that of a purely horizontal magnetic field \citep[e.g.,][]{mathis+debrye11, dhouib+22}. In reality, on the other hand, the radial component of the field is expected to dominate the interaction with the gravity waves \citep[e.g.,][]{fuller+15, mathis+21, li+22}.

Despite these difficulties, some authors have tackled the phenomenon of the magnetic mode suppression under different assumptions. 
\citet{lecoanet+17} investigated the interaction of internal gravity waves with a strong magnetic field and find perfect refraction into upward propagating slow Alfvén waves if the magnetic field strength exceeds a certain threshold. These Alfvén waves are likely to dissipate as they travel outward.
\citet{loi+papaloizou17} propose that the magneto-gravity modes might be dissipated by resonant interactions with standing torsional Alfvén modes. However, they assume that the internal magnetic field does not affect the structure of the spheroidal component of the modes, which is not the case for high field strengths \citep{loi+papaloizou18}.
In a follow-up study, \citet{loi+papaloizou18} suggest that both the toroidal and spheroidal component of the modes are affected by the magnetic field, implying that the period spacing of the suppressed dipole modes stars should be larger than that of RGB stars with normal multipole mode amplitudes.
\citet{loi20a} numerically studied the onset of the suppression by applying the Hamiltonian ray tracing method to magneto-gravity waves in the core and find that there are ray trajectories that retain a gravity-dominated character even at high magnetic field strengths.
Finally, \citet{rui+fuller23} analytically characterized the eigenfunctions of magneto-gravity modes in the presence of a dipole field and find that they do become evanescent in regions with a sufficiently strong magnetic field. However, they cannot exclude that a portion of the wave energy might escape the gravity mode cavity in a narrow equatorial band.
Apart from red-giant branch (RGB) stars, the study of \citet{lecoanet+22} suggests that the suppression of gravity modes by interaction with the magnetic field could also occur in the radiative layers of more massive main sequence (MS) stars.
A universal solution for the global oscillation modes in the presence of an internal magnetic field has yet to be found.

In this work, we investigate the transition from no mode suppression to full mode suppression using the method of \citet{loi20a}. We updated their formalism to reduce the uncertainties and calculated more sets of rays for each oscillation mode to achieve a higher resolution of the transition as a function of the magnetic field strength. In addition, we considered different stellar masses, metallicities, and topologies of the internal magnetic field. 
We subsequently used these insights into the onset of the suppression mechanism to study the evolution of the mode suppression as the star climbs the RGB.

This paper is the first in a series of publications in which we will theoretically investigate the properties of the multipole modes of stars exhibiting magnetic mode suppression where part of the mode energy entering the gravity mode cavity can escape it. Here, we present a simple expression that describes the mode suppression as a function of the magnetic field strength as predicted by our Hamiltonian ray tracing analysis. In the future, we will relate the different morphologies observed in the PSD of suppressed dipole modes stars (Coppée et al. in prep.) to physical quantities and investigate the evolution of the multipole mode amplitudes on the RGB, as well as during the helium core-burning phase.

This article is structured as follows. In Sect. \ref{sect: Ray tracing}, we describe the simulation setup of the ray tracing analysis.
In Sect. \ref{sect: Results}, we present the results of our parameter study and provide a simple expression for the magnetic mode suppression. We also study the evolution of the mode suppression along the RGB.
Finally, we discuss our findings in Sect. \ref{sect: Discussion} and summarize in Sect. \ref{sect: Summary}.

\section{Methods} \label{sect: Ray tracing}

\subsection{Stellar models} \label{sect: stellar models}

\begin{figure}[]
    \sidecaption
    \resizebox{\hsize}{!}{\includegraphics{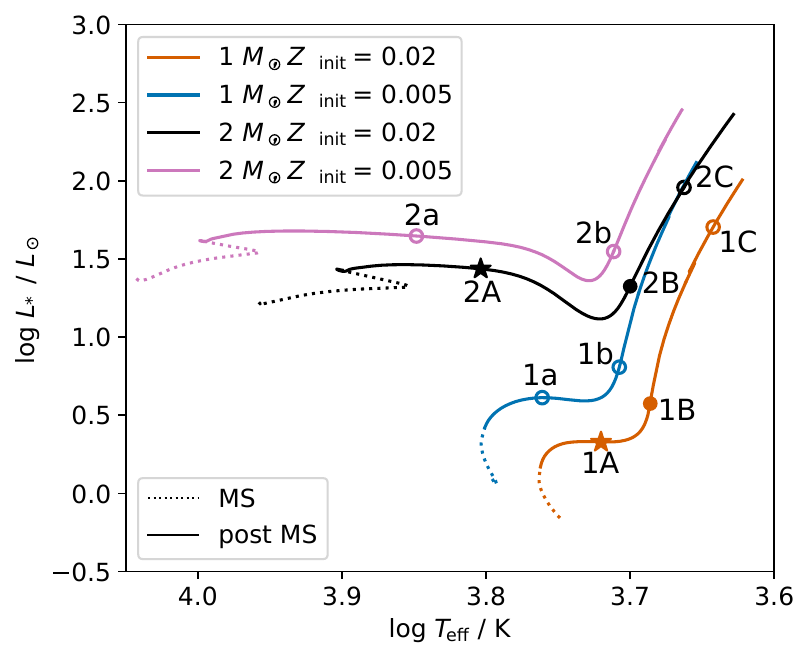}}
    \caption{Hertzsprung-Russell diagram of the four evolutionary tracks with different stellar masses and metallicities described in Sect. \ref{sect: Ray tracing}. The evolution of the MS is shown as a dotted line and the evolution after the MS as a solid line. We do not show the full evolution after the MS. The locations of the stellar models that served as the background for the ray tracing analysis are indicated by the various symbols. Star symbols indicate that the central field strength $B_{\rm cen}$, the frequency $\omega$, and the spherical degree $l$ were varied during the ray tracing analysis of a particular model. Full dots indicate that $B_{\rm cen}$ and $\omega$ were varied and circles indicate that only $B_{\rm cen}$ was varied.}
    \label{fig: HRD}
\end{figure}

\begin{table*}[]
    \caption{Selection of parameters of the stellar models shown in Fig. \ref{fig: HRD}.}
    \centering
    \begin{tabular}{c|c|c|c|c|c|c|c|c}
    \hline\hline
    Model & $M_*\ /\ M_\odot$ & $Z_{\rm init}$ & $R_*\ /\ R_\odot$ & $\rho_{\rm cen}$ / g cm$^{-3}$ & $\nu_{\rm dyn}$ / $\mu$Hz & $B_{\rm crit,cen}$ / MG & $B_{\rm crit,H}$ / MG & varied parameter(s) \\
    \hline
    1A & 1 & 0.02 & 1.8 & $9.1 \cdot 10^3$ & 42.6 & 86.2 & 36.9 & $B_{\rm cen}$, $\omega$, $l$ \\
    1B & 1 & 0.02 & 2.7 & $5.1 \cdot 10^4$ & 22.0 & 25.0 & 2.5 & $B_{\rm cen}$, $\omega$ \\
    1C & 1 & 0.02 & 12.3 & $2.2 \cdot 10^5$ & 2.3 & 0.29 & 0.006 & $B_{\rm cen}$ \\
    \hline
    1a & 1 & 0.005 & 2.0 & $6.1 \cdot 10^3$ & 34.6 & 56.6 & 30.7 & $B_{\rm cen}$ \\
    1b & 1 & 0.005 & 3.2 & $6.1 \cdot 10^4$ & 17.1 & 14.8 & 1.4 & $B_{\rm cen}$ \\
    \hline
    2A & 2 & 0.02 & 4.3 & $4.1 \cdot 10^3$ & 15.8 & 14.8 & 9.7 & $B_{\rm cen}$, $\omega$, $l$ \\
    2B & 2 & 0.02 & 6.1 & $4.4 \cdot 10^4$ & 9.4 & 4.2 & 0.7 & $B_{\rm cen}$, $\omega$ \\
    2C & 2 & 0.02 & 15.0 & $1.7 \cdot 10^5$ & 2.4 & 0.26 & 0.01 & $B_{\rm cen}$ \\
    \hline
    2a & 2 & 0.005 & 4.5 & $1.0 \cdot 10^4$ & 15.0 & 12.2 & 6.0 & $B_{\rm cen}$ \\
    2b & 2 & 0.005 & 7.5 & $5.5 \cdot 10^4$ & 6.9 & 2.2 & 0.3 & $B_{\rm cen}$ \\
    \hline\hline
    \end{tabular}
    \tablefoot{The first column gives the name of the stellar model. The second column gives the stellar mass in solar masses, the third column the stellar metallicity, the fourth column the stellar radius in solar radii, and the fifth column the central density of the star. The sixth column shows the dynamic frequency $\nu_{\rm dyn}$, which is given by $\nu_{\rm dyn} = \omega_{\rm dyn} / (2\pi)$. The seventh and eighth columns show the critical magnetic field strengths in the center of the star ($B_{\rm crit,cen}$) and in the hydrogen-burning envelope ($B_{\rm crit,H}$) for dipole modes with $\omega = 10\ \omega_{\rm dyn}$. The last column shows which parameters were varied during ray tracing analysis of a particular model. The values of models 2A, 2B, and 2C are similar to those of models A, B, and C of \citet{loi20a}, with the exception of $B_{\rm crit,cen}$ of model 2A, which is smaller than the value of model A of \citet{loi20a} by a factor of 2.}
    \label{tab: MESA models}
\end{table*}

We used version r23.05.1 of the publicly available stellar evolution code MESA \citep[Modules for Experiments in Stellar Evolution;][]{mesa1,mesa2,mesa3,mesa4,mesa5,mesa6} to compute four evolutionary tracks with varying stellar mass ($M_* = 1 M_\odot$ or $2 M_\odot$) and metallicity ($Z_{\rm init} = 0.005$ or $0.02$). We chose ten numerical models on these tracks to serve as a background for the Hamiltonian ray tracing analysis (see Sect. \ref{sect: ray initialization}). Table \ref{tab: MESA models} lists the stellar mass and metallicity of all chosen models, as well as the corresponding stellar radius $R_*$, the central density $\rho_{\rm cen}$, and the dynamical angular frequency $\omega_{\rm dyn} = \sqrt{G M_* / R_*^3}$ with $G$ being the gravitational constant.
We also show the critical magnetic field strength (Eq. \ref{eq: critical field strength}) of a dipole mode with the angular frequency $\omega = 10\ \omega_{\rm dyn}$ in the stellar center ($B_{\rm crit,cen}$) and in the hydrogen-burning shell ($B_{\rm crit,H}$). The models 2A, 2B, and 2C have the same stellar mass, metallicity, and radius as the models A, B, and C of \citet{loi20a} respectively and can be regarded as reproductions of these models. They are not identical to the models of \citet{loi20a}, because we used a different version of MESA and made small adjustments to the MESA inlist (see Appendix \ref{app: MESA inlist}). Nevertheless, they yield similar results for the ray tracing analysis. Figure \ref{fig: HRD} shows a Hertzsprung–Russell diagram of the four evolutionary tracks, in which the position of the models is indicated.  

\citet{loi20a} report some coarseness in their MESA models with a resolution similar to the grid scale. This was particularly evident in the profile of the buoyancy frequency $N$. It was likely caused by the receding convective core, which leaves small discontinuities in the profile of the chemical composition due to the finite timestep. To remove this coarseness, \citet{loi20a} smoothed the profiles of the relevant output quantities and interpolated them onto a linear grid with a lower resolution than the grid of the MESA models. Although this post-processing removed the grid-scale coarseness, the resulting profiles had a relatively small number of grid points in the core region.

Instead of performing a similar smoothing procedure, we enabled the \texttt{set\_min\_D\_mix} option in our MESA inlist and set the \texttt{min\_D\_mix} parameter to 0.5. This led to low artificial diffusive mixing everywhere in the star where the mixing was not already higher than this minimum value.
The result is a smooth profile without coarsening, while retaining the original resolution of the MESA grid. The additional mixing was not efficient enough to significantly alter the evolution of the star. In Appendix \ref{app: buoyancy frequency profiles}, we show the profile of the buoyancy frequency of the models 2A, 2B, and 2C.

\subsection{Internal magnetic field} \label{sect: magnetic field topology}

Stars with a mass $M_* \gtrsim 1.1\ M_\odot$ are assumed to have a convective core while they are on the MS. The convective motion can give rise to a dynamo, which can generate a magnetic field. When the star evolves away from the MS and becomes a subgiant and later a red giant, the core is fully radiative and the potential dynamo can no longer operate. However, the magnetic field generated during the dynamo phase might stabilize into a large-scale magnetic field \citep[e.g.,][]{braithwaite08, emeriau-viard+17, villebrun+19}. Since the magnetic diffusion timescale has been estimated to be small for magnetic fields contained in the radiative core \citep[e.g.,][]{cantiello+16}, they could survive the post-MS evolution. 
The strength of these buried magnetic fields is expected to be relatively large on the RGB. The reason for this is that the radiative core harboring the magnetic fields contracts after the MS, which increases the field strength due to the conservation of magnetic flux \citep[see][]{fuller+15, cantiello+16, bugnet+21}.

Both analytical and numerical studies have shown that purely toroidal or poloidal magnetic fields are unstable \citep[e.g.,][]{taylor73, markey+taylor73, braithwaite06, braithwaite07}. Instead, stable magnetic field configurations are expected to have both toriodal and poloidal components of significant strength \citep[e.g.,][]{taylor80, braithwaite08, akgun+13, becerra+22} and to take the form of a twisted torus configuration \citep{braithwaite+nordlund06, braithwaite+spruit17}. 
A commonly used analytical prescription for an equilibrium axisymmetric twisted torus field was derived by \citet{prendergast56} and later generalized by \citet{duez+mathis10} to the compressible case. This formalism is sometimes referred to as the "Prendergast field" and has several useful properties. First, the magnetic field can be easily constructed from analytic expressions that can be directly applied to stellar models. Second, the field resembles a dipole field, but also has a significant toroidal component. 
Third, the field strength does not diverge at the center of the star and has a finite maximum value. Finally, all components of the field vanish as the distance from the center approaches a pre-selected maximum radius.
Note that although Prendergast fields satisfy the stability criterion established by various studies \citep[e.g.,][]{braithwaite09, akgun+13, becerra+22} in stably stratified stellar models,
recent findings by \citet{kaufman+22} indicate that these fields may not be stable throughout stellar evolution. For our purposes, this is not crucial because we demonstrate in Sect. \ref{sect: applicability f_T} that our results are expected to be valid for any field topology with a dipole-like dependence on latitude and a radial component comparable or larger than the latitudinal and longitudinal component.

The Prendergast field is described in terms of the radial flux function $\Psi(r)$ given by
\begin{gather}
    \Psi(r) = \frac{\beta\lambda r}{j_1(\lambda r_{\rm f})}  {\Bigg [} f_\lambda(r,r_{\rm f}) \int_0^r \rho \xi^3 j_1(\lambda\xi) {\rm d}\xi \notag\\
        \qquad\qquad\qquad\qquad + j_1(\lambda r) \int_r^{r_{\rm f}} \rho\xi^3 f_\lambda(\xi,r_{\rm f}) {\rm d}\xi {\Bigg ]},
\end{gather}
with $f_\lambda(r_1, r_2) \equiv j_1(\lambda r_2) y_1(\lambda r_1) - j_1(\lambda r_1) y_1(\lambda r_2)$.
Here, $r$ is the distance to the center of the star, $r_{\rm f}$ is the maximum radial extend of the magnetic field, $\rho$ is the density of the star, $\beta$ is a scaling factor that controls the field strength, $j_1$ and $y_1$ are spherical Bessel functions, and $\lambda$ can be determined by
\begin{gather}
    \int_0^{r_{\rm f}} \rho\xi^3 j_1(\lambda\xi) {\rm d}\xi = 0.
    \label{eq: lambda requirement}
\end{gather}
Finding values of $\lambda$ that satisfy Eq. \ref{eq: lambda requirement} is usually possible for subgiant and red giant models for a given $r_{\rm f}$. In this work, we always chose the smallest possible value of $\lambda$ for each stellar model. The value for $r_{\rm f}$ used for a model can be found in Table \ref{tab: model initialization}. 
The components of the magnetic field $\boldsymbol{\rm B} = (B_r, B_\theta, B_\phi)$ are given by
\begin{gather}
    B_{\rm r} = \frac{2}{r^2} \Psi(r) \cos\theta, \label{eq: B_r}\\
    B_{\theta} = - \frac{1}{r} \Psi^\prime(r) \sin\theta, \\
    B_\phi = - \frac{\lambda}{r} \Psi(r) \sin\theta,
\end{gather}
which we used to insert a magnetic field into the radiative core of each of our MESA models before performing the ray tracing analysis.

The strength of the components of the magnetic field is scaled by adjusting $\beta$ such that the absolute value of the component with the highest central field strength is equal to a preselected parameter $B_{\rm cen}$ at the center of the star. We refer to $B_{\rm cen}$ as the central field strength. It is specified in terms of the critical field strength
\begin{gather}
    B_{\rm crit} = \sqrt{\mu_0 \rho} \frac{r}{N} \frac{\omega^2}{\sqrt{l(l+1)}},
    \label{eq: critical field strength}
\end{gather}
where $l$ is the spherical degree of the oscillation mode and $\mu_0$ is the magnetic permeability. It marks the approximate field strength at which the magnetic field is strong enough to be dynamically relevant for the oscillations\footnote{Note that this definition of the critical field strength is different by a factor of 2 from the one used by \citet{fuller+15}.}. In this work, we always express the field strength in the center $B_{\rm cen}$ as a factor that is multiplied by the central critical field strength $B_{\rm crit,cen}$.
The internal magnetic field strengths considered in this work are not high enough to significantly influence the evolution of the star, which means that the magnetic effects can be neglected in the MESA computations.

\subsection{Hamilton's equations} \label{sect: hamilton}

Hamiltonian ray tracing describes the propagation of a group velocity wave packet by assigning a position $\boldsymbol{\rm x} = (r,\theta,\phi)$ and a wavevector $\boldsymbol{\rm k} = (k_r,k_\theta,k_\phi)$ to the wave packet at any time $t$. The trajectory of the wave packet is described by Hamilton's equations:
\begin{gather}
    \frac{{\rm d}\boldsymbol{\rm x}}{{\rm d}t} = \nabla_{\boldsymbol{\rm k}} H, \quad \frac{{\rm d}\boldsymbol{\rm k}}{{\rm d}t} = - \nabla H,
\end{gather}
where the Hamiltonian $H$ is given by the dispersion relation $\omega$ (see Eq. \ref{eq: dispersion relation}). In this work, we focus on the stellar core region, where the sound speed is so high that the dynamics of the oscillations are dominated by buoyancy and the interaction with a potential internal magnetic field (i.e., magneto-gravity waves). The dispersion relation of magneto-gravity waves is:
\begin{gather}
    \omega^2 = \omega_{\rm A}^2 + \omega_{\rm g}^2 = (\boldsymbol{\rm k} \cdot {\boldsymbol{\rm V}}_{\rm A})^2 + \left(\frac{k_\perp}{k} N \right)^2.
    \label{eq: dispersion relation}
\end{gather}
Here, $\omega_{\rm A}$ is the Alfvén wave frequency, $\omega_{\rm g}$ is the gravity wave frequency, ${\boldsymbol{\rm V}}_{\rm A} = (V_{{\rm A}r},V_{{\rm A}\theta},V_{{\rm A}\phi})$ is the Alfvén velocity, $k = \sqrt{k^2_r + k^2_\theta + k^2_\phi}$ is the wavenumber, and $k_\perp = \sqrt{k^2_\theta + k^2_\phi}$ is the horizontal wavenumber.

\citet{loi20a} formulated the Hamilton's equations in the case of magneto-gravity waves in three-dimensional spherical polar coordinates, resulting in a system of six linear first-order ordinary differential equations. The derivation can be found in \citet{loi20a}. For completeness, we repeat the equations in Appendix \ref{app: Equations Hamiltonian} (Eqs. \ref{eq: hamiltion r} to \ref{eq: hamilton k_phi}).

It is important to note that this approach relies on the Wentzel–Kramers–Brillouin–Jeffreys (WKBJ) approximation in both the radial and horizontal direction, which requires that the wavelength is smaller than the length scale of the variation of the stellar structure. For non-magnetized red giant stars, this is typically a good assumption. However, the picture becomes more complicated when an internal magnetic field is present, since the horizontal wavenumber of the observable modes (i.e., modes with low $l$) has a length scale comparable to that of the horizontal variation of the magnetic field. The resulting limitations and our motivation for nonetheless using this approach are described in Sect. \ref{sect: limitations}. Overall, we do not expect a strong impact on our main results.

\subsection{Ray initialization and integration} \label{sect: ray initialization}

\begin{table}[]
    \caption{Initialization parameters of the ray tracing analysis for the various stellar models.}
    \centering
    \begin{tabular}{c|c|c|c}
    \hline\hline
    Model & $r_0\ /\ R_{*}$ & $r_{\rm f}\ /\ R_{*}$ & $t_{\rm int}\ /\ \omega_{\rm dyn}^{-1}$ \\
    \hline
    1A & 0.06 & 0.024 & 50 \\
    1B & 0.02 & 0.011 & 200 \\
    1C & 0.0021 & 0.0016 & 2000 \\
    \hline
    1a & 0.06 & 0.024 & 50 \\
    1b & 0.02 & 0.011 & 200 \\
    \hline
    2A & 0.04 & 0.016 & 100 \\
    2B & 0.01 & 0.0055 & 400 \\
    2C & 0.0021 & 0.0016 & 2000 \\
    \hline
    2a & 0.04 & 0.016 & 100 \\
    2b & 0.01 & 0.0055 & 400 \\
    \hline\hline
    \end{tabular}
    \tablefoot{The meaning of the parameters is described in the main text (see Sects. \ref{sect: magnetic field topology} and \ref{sect: ray initialization}).}
    \label{tab: model initialization}
\end{table}

We solved the magneto-gravity ray tracing equations (Eqs. \ref{eq: hamiltion r} to \ref{eq: hamilton k_phi}) by integrating them forward in time using a fourth-order Runge-Kutta scheme. The integration timestep was always set to $0.001\ \omega_{\rm dyn}^{-1}$. The total integration time $t_{\rm int}$ of each model is listed in Table \ref{tab: model initialization}. To initialize a ray, we selected a numerical stellar model which serves as the background.
In addition, we chose the frequency of the wave packet $\omega$, the spherical degree $l$, and the initial latitude $\theta_0$.
We also specified a polarization angle $\alpha$ with $0 \leq \alpha \leq 2\pi$, which defines the initial horizontal orientation of the ray.

Using these quantities, we determined the starting values for the six parameters to be integrated (i.e., $r_0$, $\theta_0$, $\phi_0$, $k_{r0}$, $k_{\theta 0}$, $k_{\phi 0}$) in the same way as \citet{loi20a}.
The exact value of the initial radius $r_0$ does not matter as long as it is inside the radiative core and exterior to the maximum extend of the internal magnetic field $r_{\rm f}$ \citep{loi20a}. Therefore, we did not test different values of $r_{\rm 0}$ for a given model. The value for $r_0$ used for each model is listed in Table \ref{tab: model initialization}. 
The initial latitude $\theta_0$ can take values from 0 to $\pi$ and the choice of the initial longitude $\phi_0$ is irrelevant in the context of this work, since we only considered axisymmetric magnetic fields. We followed \citet{loi20a} and set $\phi_0 = \alpha$. The starting values of the components of the wavevector are given by
\begin{gather}
    k_{r0} = \sqrt{k^2_0 - k^2_{\perp 0}}, \\
    k_{\theta 0} = k_{\perp 0} \sin\alpha, \\
    k_{\phi 0} = k_{\perp 0} \cos\alpha,
\end{gather}
with $k_{\perp 0} = \sqrt{l(l+1)/r_0}$ and $k_0 = k_{\perp 0} N(r_0)/\omega$.

We approximate the behavior of a global oscillation mode by calculating a set of 1200 individual rays and subsequently averaging over the initial latitude $\theta_0$ and the polarization angle $\alpha$. 
This is intended to mimic the arbitrariness of the propagation direction of waves traveling on a spherically symmetric background.
Each set of rays is characterized by the parameters $\omega$, $l$, and $B_{\rm cen}$, and is assigned to a specific stellar model that serves as the background. 
The values used when varying frequency, the spherical degree, and the magnetic field strength are $\omega = (8, {\bf 10}, 12)\ \omega_{\rm dyn}$, $l = ({\bf 1}, 2, 3)$, and $B_{\rm cen} = (0.1, 0.6, 0.7, 0.8, 1, 2, 3, 10)\ B_{\rm crit,cen}$, where the values of $\omega$ and $l$, which are used if they are not varied, are shown in bold.
We chose $\theta_0$ and $\alpha$ in such a way that the initial positions of the rays are evenly distributed on the surface of a sphere when $\theta_0$ is considered as the latitude and $\alpha$ as the longitude.

The Runge-Kutta integration scheme does not inherently conserve $\omega$ \citep{loi20a}\footnote{Integration methods that inherently conserve $\omega$ did face convergence issues and were therefore not used \citep{loi20a}.}. However, since the Hamiltonian has no explicit time dependence, it is crucial that $\omega$ remains constant throughout the integration. While the integration routine conserves $\omega$ to machine precision when using infinitely differentiable background profiles \citep{loi20a}, the finite grid of the MESA models introduces errors that accumulate and render the resulting trajectories of the rays unusable for further analysis. \citet{loi20a} attempted to correct these errors by adjusting the value of $k_r$ after each integration timestep such that $\omega$ remains constant. This was done by performing a Newton–Raphson iteration scheme for a maximum of five iterations to solve Eq. \ref{eq: fix k_r 1} (see Appendix \ref{app: adjustment wavevector}). Although this correction ensured the conservation of $\omega$ within 1 \% of the original value for most of their rays, the Newton-Raphson scheme did sometimes not converge, leading to uncertainties on the behavior of the global mode.

Here, we built on the method of \citet{loi20a} and also adjusted $k_r$ after each integration timestep to ensure that $\omega$ remains constant.
However, instead of using a Newton–Raphson scheme, we directly calculated the roots of Eq. \ref{eq: fix k_r 1} after each timestep. We therefore faced no convergence problems.
The roots can be complex. We then identified the root with the real part closest to our current estimate of $k_r$. The imaginary part of the wavenumber is typically small and corresponds to the damping or excitation of the oscillations, which are not included in the ray tracing formalism. 
We thus changed the current value of $k_r$ to the real part of the selected root after each timestep. If the imaginary part was unequal to zero, Eq. \ref{eq: fix k_r 1} was not satisfied anymore and $\omega$ was not conserved.
The possible discrepancy between the initial value of $\omega$ and the value of $\omega$ after each timestep depends on how much the imaginary part of the selected root deviated from 0.
Therefore, the value of $\omega$ after each timestep is a sensitive measure of accuracy of the rays. 
If the value of $\omega$ was different to the original value by more than 0.0001 \% at any point during the integration, we did not consider that ray for further analysis\footnote{The threshold value for the relative change in $\omega$ was chosen so that we can reliably filter out rays with trajectories that exhibit clear numerical artifacts.}. 
We compare our method for correcting $k_r$ and the Newton–Raphson scheme used by \citet{loi20a} in Appendix \ref{sect: setup comparison}.

\subsection{Trapped fraction} \label{sect: trapped fraction}

\begin{figure*}[]
    \sidecaption
    \resizebox{\hsize}{!}{\includegraphics{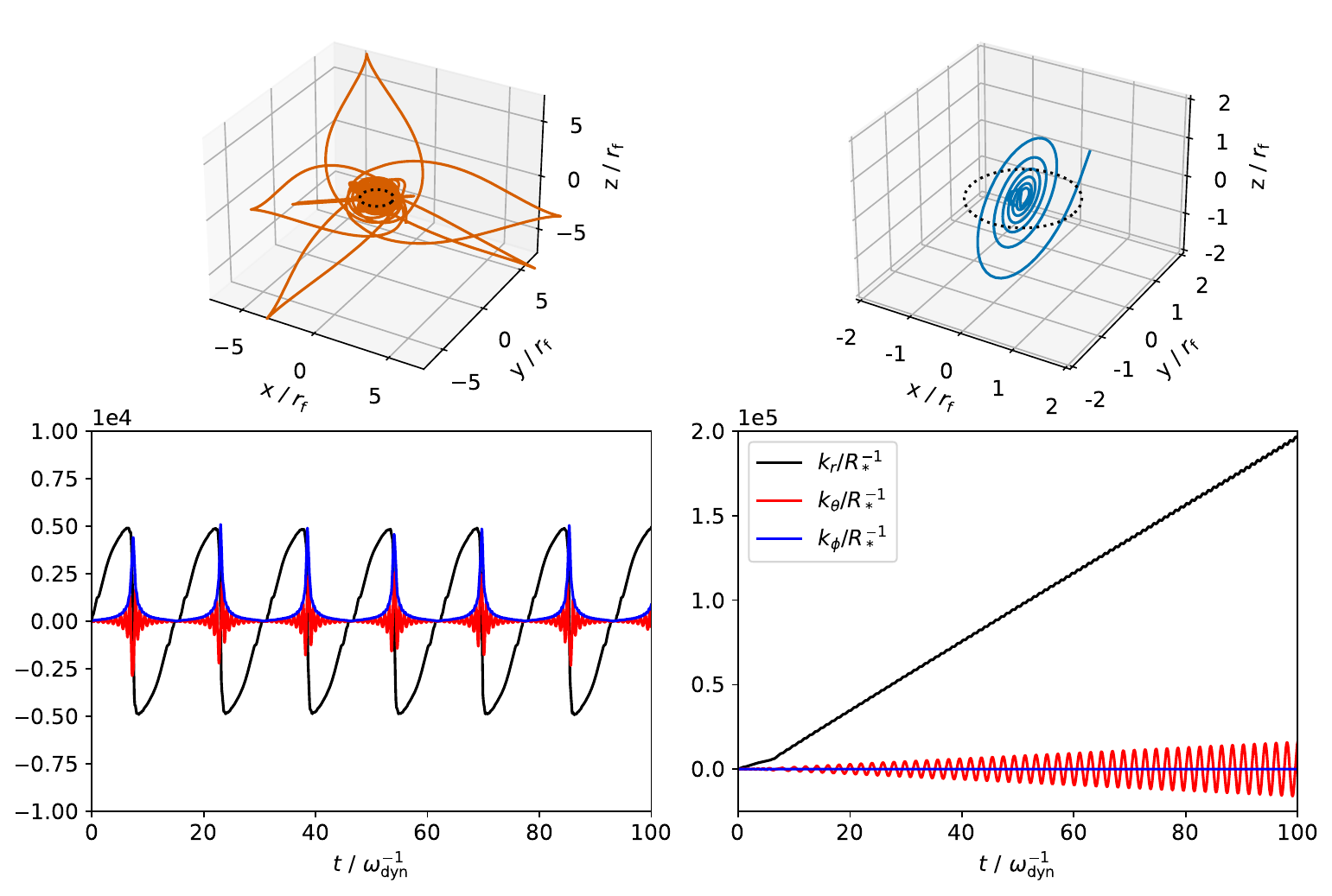}}
    \caption{Trajectory of a reflected and trapped ray \textit{(top row)} and the corresponding wavevector components as a function of time \textit{(bottom row)} for the model 2A with $B_{\rm cen} = B_{\rm crit,cen}$, $\omega = 10\ \omega_{\rm dyn}$, and $l = 1$. The reflected ray with $\theta_0 = 128.9^{\circ}$ and $\alpha = 3.7^{\circ}$ is shown on the left, the trapped ray with $\theta_0 = 98.7^{\circ}$ and $\alpha = 90.9^{\circ}$ is shown on the right. The dotted black circle in the top row represents the maximum extend of the magnetic field $r_{\rm f}$. Note that the limits of the axes do not coincide between left and right for the sake of clarity.}
    \label{fig: reflected and trapped rays}
\end{figure*}

There are two possible outcomes for rays launched into the magnetized stellar core: They can be either reflected or trapped \citep{loi20a}. In Fig. \ref{fig: reflected and trapped rays}, we show examples for both a reflected and a trapped ray.
Reflected rays propagate back and forth (in distance from the center) between two turning points and regularly enter and exit the area of the internal magnetic field. As shown in Fig. \ref{fig: reflected and trapped rays}, the components of the wavevector, and thus also the wavenumber of a reflected ray, undergo periodic changes, but remain bounded in time. This is in contrast to the trapped rays. Once a trapped ray enters the area of the magnetic field, it never emerges and the wavenumber diverges. Since the rays visualize the paths of the energy flow of the oscillations, a trapped ray can be interpreted as a loss of the observable mode energy. 
To decide whether a ray is reflected or trapped, we used a classification scheme, which is described in Appendix \ref{app: ray classification}. It is an updated version of the ray classification scheme used by \citet{loi20a}. 
We compare our ray classification scheme with that of \citet{loi20a} in Appendix \ref{sect: setup comparison}.

When all 1200 rays of a set have been classified as either reflected or trapped, or identified as unclassified due to numerical issues during the integration (see Appendix \ref{app: ray classification}), the trapped fraction $f_{\rm T}$ of that set of rays could be determined. It is an estimate of the fraction of energy in the gravity mode cavity that is lost from the observable global mode due to the interaction with the internal magnetic field. The trapped fraction can be calculated as follows:
\begin{gather}
    f_{\rm T} = \frac{1}{{\rm \#total}} \left( {\rm \#trapped} + \frac{{\rm \#unclassified}}{2} \pm \frac{{\rm \#unclassified}}{2} \right),
    \label{eq: trapped fraction}
\end{gather}
where ${\rm \#trapped}$ is the number of trapped rays, ${\rm \#unclassified}$ is the number of unclassified rays, and ${\rm \#total} = 1200$ is the total number of rays in the set. 
The estimated value of $f_{\rm T}$ corresponds to the case where half of the unclassified rays are actually reflected and the other half is trapped.
The uncertainty of $f_{\rm T}$ introduced by the last term in Eq. \ref{eq: trapped fraction} is numerical and corresponds to the fact that not all rays of a set could be identified as reflected or trapped. If these numerical difficulties did not exist, we would expect the unclassified rays to be either reflected or trapped. The minimum value of $f_{\rm T}$ permitted by the uncertainties therefore represents the case in which all unclassified rays are reflected, while the maximum value of $f_{\rm T}$ represents the case in which all unclassified rays are trapped.

\section{Results} \label{sect: Results}

Before we discuss the results of our magneto-gravity ray tracing analysis, we summarize the changes that we have made to the procedure of \citet{loi20a} described in Sect. \ref{sect: Ray tracing}. First, we used input stellar structure profiles with an improved radial resolution. This was achieved by smoothing the profiles directly in MESA by introducing a small amount of additional mixing (see Sect. \ref{sect: stellar models}).
Second, we chose the initial positions of the rays so that they are evenly distributed on the surface of a sphere if $\theta_0$ is considered as latitude and $\alpha$ as longitude (see Fig. \ref{fig: unispheres newton}).
Third, we used a different method to correct the wavenumber after each timestep during the integration (see Sect. \ref{sect: ray initialization} and Appendix \ref{app: adjustment wavevector}). Finally, we used an updated ray classification scheme (see Appendix \ref{app: ray classification}).

\subsection{Numerical estimates of the trapped fraction} \label{sect: results ray tracing}

\begin{figure}[]
    \sidecaption
    \resizebox{\hsize}{!}{\includegraphics{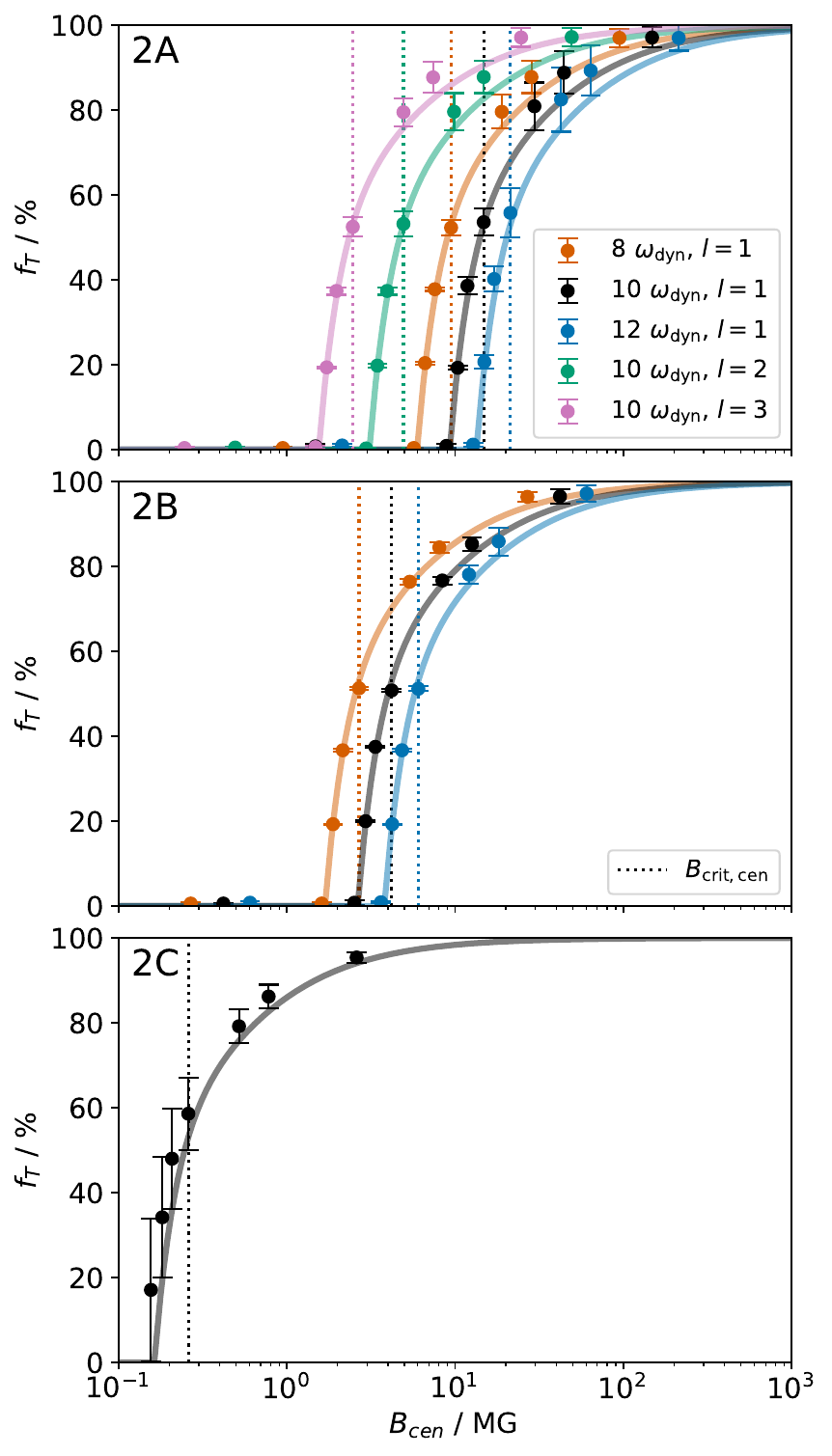}}
    \caption{Trapped fraction as a function of the central field strength of the internal magnetic field for the models 2A \textit{(top)}, 2B \textit{(middle)}, and 2C \textit{(bottom)}. Different colors indicate different combinations of the frequency $\omega$ and the spherical degree $l$ (see legend). The dotted lines mark the corresponding central critical field strength. The error bars are determined by the uncertainty of $f_{\rm T}$ (see Eq. \ref{eq: trapped fraction}). The line connecting the symbols is a fit to the trapped fraction presented in Sect. \ref{sect: analytic trapped fraction}.}
    \label{fig: trapped fraction model 2}
\end{figure}

\begin{figure}[]
    \sidecaption
    \resizebox{\hsize}{!}{\includegraphics{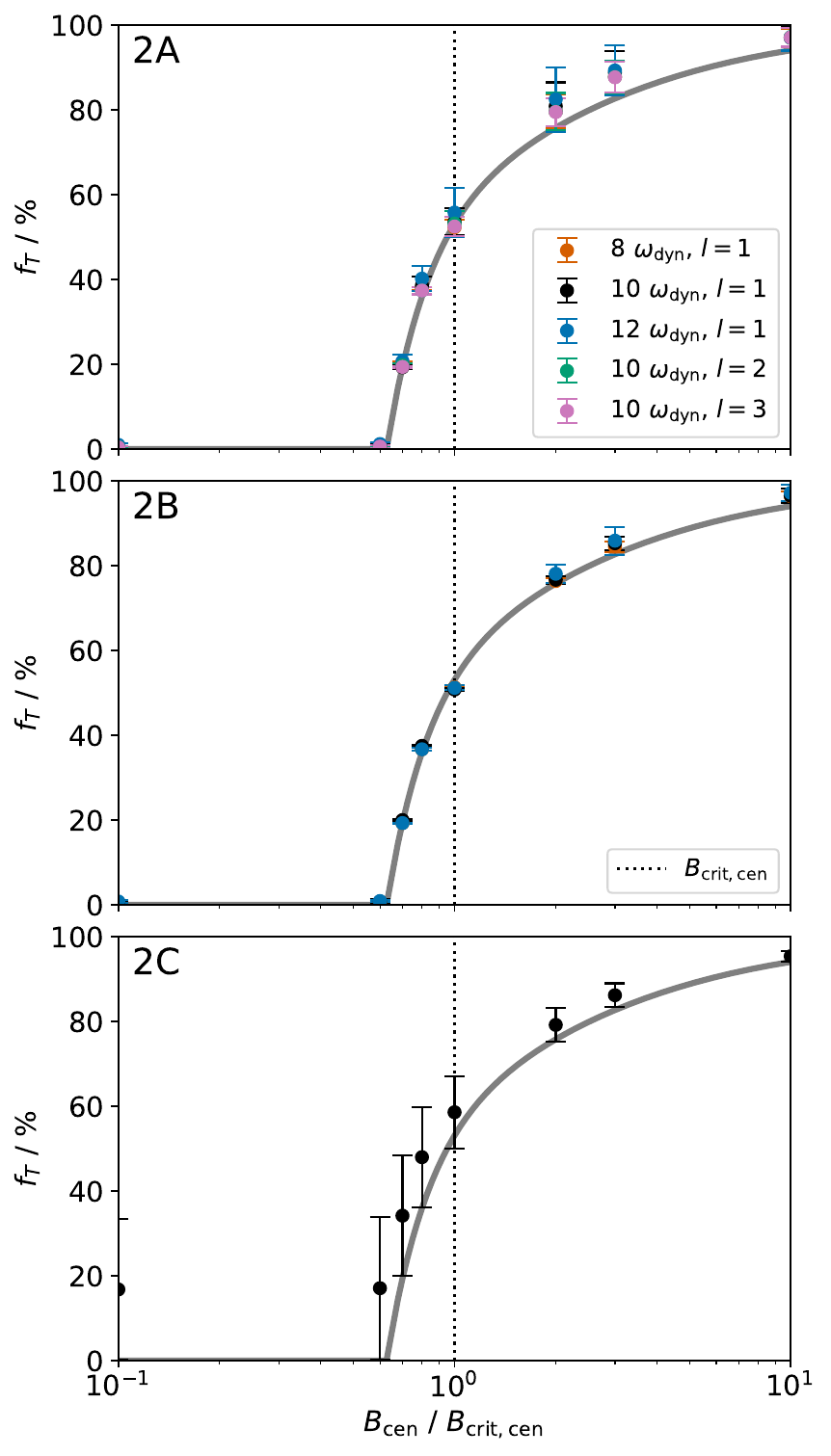}}
    \caption{Same as Fig. \ref{fig: trapped fraction model 2}, now as a function of the central field strength of the internal magnetic field divided by the corresponding central critical field strength. The upper limit of the horizontal axis is now 10. Note that $B_{\rm crit,cen}$ is different for each model, as well as for each combination of $\omega$ and $l$ (see Eq. \ref{eq: critical field strength}).}
    \label{fig: trapped fraction model 2 Bcrit}
\end{figure}

In Fig. \ref{fig: trapped fraction model 2}, we show the trapped fraction of the models 2A, 2B, and 2C as a function of the magnetic field strength $B_{\rm cen}$ for different combinations of $\omega$ and $l$.
Although we used a similar method as \citet{loi20a} to compute $f_{\rm T}$, the trapped fractions calculated in this work are smaller than those obtained by \citet{loi20a} when comparing our model 2A (2B, 2C) to their model A (B, C). The reason for this is most likely a systematic difference in the propagation of the rays. We discuss these differences in Appendix \ref{app: comparison to loi}. In particular, we show that the trajectories of our rays match those of \citet{loi20b}, who also uses a slightly adapted version of the ray integration procedure presented in \citet{loi20a}.

Overall, we recover three qualitative results reported by \citet{loi20a}:
\begin{itemize}
    \item The trapped fraction $f_{\rm T}$ is negligible when $B_{\rm cen}$ is smaller than a certain threshold value. If $B_{\rm cen}$ is greater than this threshold, $f_{\rm T}$ increases with increasing $B_{\rm cen}$.
    \item The threshold value of $B_{\rm cen}$ is lower for lower $\omega$. If $B_{\rm cen}$ is greater than the threshold, $f_{\rm T}$ is higher at a certain $B_{\rm cen}$ when $\omega$ is lower. 
    \item The threshold value of $B_{\rm cen}$ is lower for higher $l$. If $B_{\rm cen}$ is greater than the threshold, $f_{\rm T}$ is higher at a certain $B_{\rm cen}$ when $l$ is higher. 
\end{itemize}
Furthermore, we find that $f_{\rm T}$ always approaches 100 \% for values of $B_{\rm cen}$ much higher than the corresponding central critical field strength $B_{\rm crit,cen}$. This applies to all models shown in Fig. \ref{fig: trapped fraction model 2}, including the more evolved models 2B and 2C, and is in contrast to the trend with stellar evolution observed by \citet{loi20a}. They conclude that the maximum value of $f_{\rm T}$ appears to saturate at values less than 100 \% for their more evolved models. This discrepancy cannot be explained by the uncertainties of their $f_{\rm T}$-estimate. However, as we show in Appendix \ref{sect: setup comparison}, it is possible that these uncertainties are underestimated due to numerical artifacts. This means that the value at which $f_{\rm T}$ saturates could actually be consistent with 100 \% also for their more evolved models.

As shown in Fig. \ref{fig: trapped fraction model 2}, the exact value of $f_{\rm T}$ as a function of $B_{\rm cen}$ depends on the input stellar model, $\omega$, and $l$. The same can be said about $B_{\rm crit,cen}$, which also depends on these parameters (see Eq. \ref{eq: critical field strength}). In Fig. \ref{fig: trapped fraction model 2 Bcrit}, we show $f_{\rm T}$ for the same models as in Fig. \ref{fig: trapped fraction model 2} with the central magnetic field strength expressed in terms of $B_{\rm crit,cen}$. In short, this means that we aligned the dotted lines indicating $B_{\rm crit,cen}$ shown in Fig. \ref{fig: trapped fraction model 2}. Figure \ref{fig: trapped fraction model 2 Bcrit} shows that the entire dependence of $f_{\rm T}$ on the stellar model, $\omega$, and $l$ is explained by the dependence of $B_{\rm crit,cen}$ on these parameters. Crucially, this means that $f_{\rm T}$ can be expressed as a function of $B_{\rm cen}/B_{\rm crit,cen}$ only. Therefore, $f_{\rm T}$ can be directly estimated for a given $B_{\rm cen}$ if $B_{\rm crit,cen}$ is known.

This also holds true if the mass and metallicity of the star are changed. We show plots similar to Fig. \ref{fig: trapped fraction model 2 Bcrit} for the models 1A, 1B, 1C, 1a, 1b, 2a, and 2b in Appendix \ref{app: additional plot ray tracing}.

\subsection{Fitted expression for the trapped fraction} \label{sect: analytic trapped fraction}

The trapped fraction can be expressed as a function of the ratio between central field strength and the central critical field strength (see Fig. \ref{fig: trapped fraction model 2 Bcrit}). It is therefore possible to fit an analytical expression to $f_{\rm T}$ that only depends on $B_{\rm cen}/B_{\rm crit,cen}$. Since $B_{\rm crit,cen}$ can be determined using Eq. \ref{eq: critical field strength}, $f_{\rm T}$ is reduced to a function of $B_{\rm cen}$ for a given stellar model.

For simplicity, we only considered the model 2b with $\omega = 10\ \omega_{\rm dyn}$ and $l = 1$ for the fitting procedure. We chose this combination of parameters because it yields small uncertainties for $f_{\rm T}$ and shows good agreement with the trapped fractions of the other models.
The resulting fitted function describing the trapped fraction is as follows:
\begin{gather}
    f_{\rm T}^{\rm (fit)}(B) = \max \left[ \ 0,\ 1 - 0.635\ \left( \frac{B}{B_{\rm crit}} \right)^{-1} \right. \notag\\
    \qquad\qquad\qquad \left. + 0.438\ \left( \frac{B}{B_{\rm crit}} \right)^{-2} - 0.272\ \left( \frac{B}{B_{\rm crit}} \right)^{-3} \right].
    \label{eq: trapped fraction fit}
\end{gather}
According to this expression, the onset of the magnetic suppression takes place at $B \approx 0.628\ B_{\rm crit}$.
Here, we have omitted the subscript $_{\rm cen}$ because we show in the following section that Eq. \ref{eq: trapped fraction fit} is also valid when considering the ratio between the field strength and the critical field strength at a location other than the center of the star.
Note that the choice of the form of the function describing $f_{\rm T}^{\rm (fit)}$ is purely empirical.

The fit to the trapped fraction is shown in Figs. \ref{fig: trapped fraction model 2} and \ref{fig: trapped fraction model 2 Bcrit}, as well as Figs. \ref{fig: trapped fraction model 1 Bcrit} and \ref{fig: trapped fraction model lowZ Bcrit}, together with the numerical results from the ray tracing analysis. We discuss the applicability of Eq. \ref{eq: trapped fraction fit} in the following section.

\subsection{Applicability of the fitted trapped fraction} \label{sect: applicability f_T}

\begin{figure}[]
    \sidecaption
    \resizebox{\hsize}{!}{\includegraphics{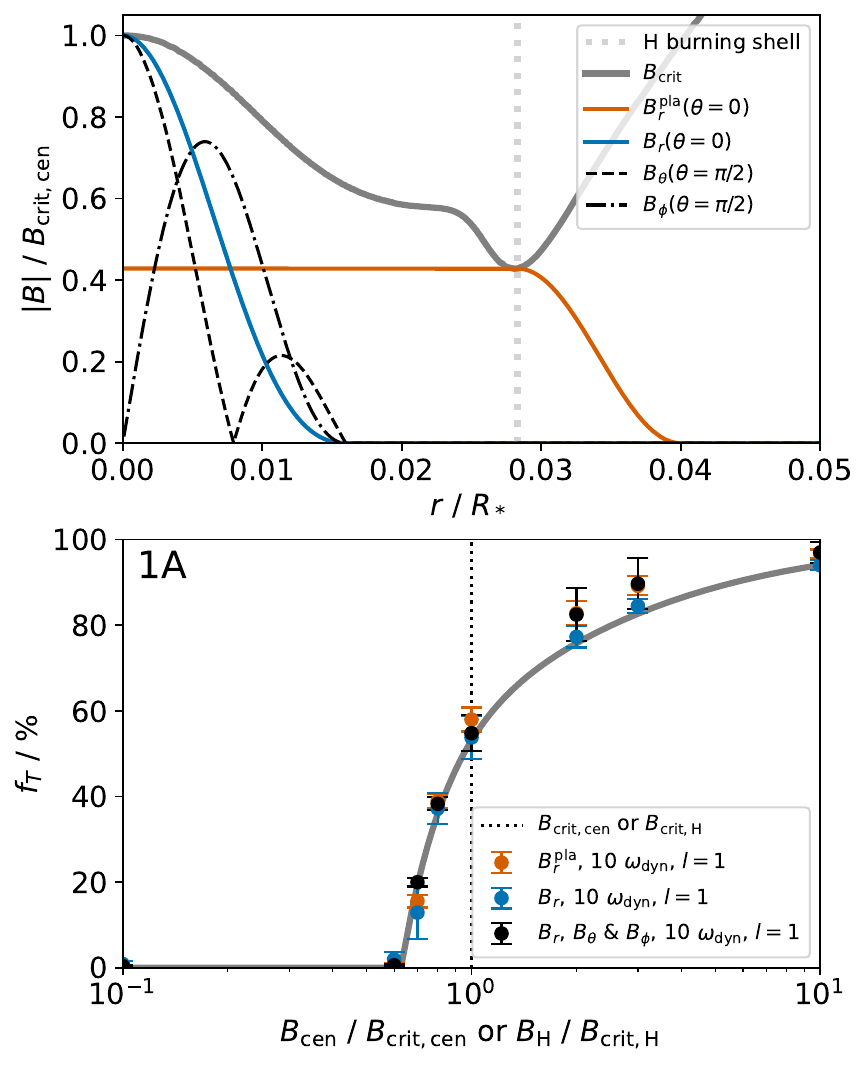}}
    \caption{{\it Top:} Normalized critical field strength as a function of radius. We also show the maximum absolute value of the three components of the Prendergast field (see Sect. \ref{sect: magnetic field topology}), which is scaled to $B_{\rm cen} = B_{\rm crit,cen}$. Furthermore, we show the radial component of an improvised magnetic field that has a plateau and scales with the critical field strength at the hydrogen-burning shell $B_{\rm crit,H}$. The strength of the improvised field at the hydrogen-burning shell is set to $B_{\rm H} = B_{\rm crit,H}$. The hydrogen-burning shell is indicated by the vertical dotted line. {\it Bottom:} Trapped fraction as a function of the central field strength or the field strength at the hydrogen-burning shell for model 1A with $\omega = 10\ \omega_{\rm dyn}$ and $l=1$. The black symbols correspond to the case where all components of the Prendergast field are taken into account. They are the same as the black symbols in the top panel of Fig. \ref{fig: trapped fraction model 1 Bcrit}. The blue symbols indicate that only the radial component of the Prendergast field has been used and the other components were set to zero. The red symbols correspond to magnetic field featuring the plateau that is scaled by $B_{\rm crit,H}$ instead of $B_{\rm crit,cen}$. The dotted line marks either $B_{\rm crit,cen}$ or $B_{\rm crit,H}$. The line connecting the symbols is the fit to the trapped fraction (Eq. \ref{eq: trapped fraction fit}).}
    \label{fig: H bruning shell}
\end{figure}

Since we have only considered Prendergast field configurations (see Sect. \ref{sect: magnetic field topology}) for the calculation of the trajectories of the rays, Eq. \ref{eq: trapped fraction fit} is strictly speaking only valid for exactly this type of internal magnetic field. Adopting other magnetic field topologies, such as a pure dipole field, led to numerical difficulties and likely requires changes to our integration method. This will be subject of a future work. However, we can deduce some characteristics of how the behavior of the rays is expected to change with a modification of the field topology from the results presented in this work.

In Fig. \ref{fig: H bruning shell}, we compare the trapped fraction obtained when using a Prendergast field with the trapped fraction when only the radial component of the Prendergast field is considered (i.e., $B_\theta = B_\phi = 0$), as well as with the radial component of an improvised magnetic field $B_r^{\rm pla}$. The improvised field has a plateau and scales with the critical field strength at the hydrogen-burning shell $B_{\rm crit,H}$. 
In addition, it has a dipole-like dependence on the latitude, similar to the Prendergast field (i.e., $B_r^{\rm pla} \propto \cos\theta$). For the Prendergast field, Fig. \ref{fig: H bruning shell} shows that $f_{\rm T}$ does not change significantly when neglecting $B_\theta$ and $B_\phi$, even though $B_\theta$ has the same maximum absolute strength as $B_r$. This indicates that $B_r$ dominates the interaction of the magnetic field with the rays when the absolute values of $B_\theta$ and $B_\phi$ are roughly comparable or smaller than that of $B_r$. 

Interestingly, the behavior of the trapped fraction as a function of  field strength when considering the improvised magnetic field does not differ significantly from the case where the Prendergast field is used. This applies under the condition that the improvised field is scaled by the field strength at the hydrogen-burning shell $B_{\rm H} / B_{\rm crit,H}$ instead of $B_{\rm cen} / B_{\rm crit,cen}$. It indicates that it does not matter where in the core the field strength first approaches the critical field strength, and that the ratio between the actual field strength and the critical field strength at that location determines $f_{\rm T}$.
A prime location for this to happen is the hydrogen-burning shell, as the critical field strength is at a minimum there.
In our models, the Prendergast field always reaches the critical field strength in the center of the star first, even if $r_{\rm f}$ lies outside the hydrogen-burning shell.
Since the critical field strength at the center and at the hydrogen-burning shell differ by up to almost two orders of magnitude for some models (see Table \ref{tab: MESA models}), different field topologies that approach the critical field strength at either the center or the hydrogen-burning shell could cause a dichotomy in the trapped fraction.

When representing a set of rays as a unisphere parameterized by $\theta_0$ and $\alpha$ (see Fig. \ref{fig: unispheres newton}), we can see a clear structure in the distribution of reflected and trapped rays. The reflected rays are localized at two poles that are located opposite to each other. If the trapped fraction of the set of rays increases, the poles of reflected rays become smaller; if the trapped fraction decreases, the poles expand. The reflected rays are forming these poles because they share two properties: A small initial launch latitude $\theta_0$ and a small latitudinal component of the wavevector $k_\theta$.
This means that the reflected rays are those rays that propagate in an equatorial band. The width of this band is being controlled by $B_{\rm cen}$. This is related to the fact that the radial component of the magnetic field $B_r$ is proportional to $\cos \theta$ (Eq. \ref{eq: B_r}), which means that $B_r$ vanishes at the equator. Since this is a characteristic of dipole fields, the same trend is likely to be observed for all dipole-type field topologies. However, it is not the case for quadrupole fields or other higher-order topologies. Therefore, we expect the rays to behave differently for these configurations\footnote{Different field configurations are expected to change the distribution of reflected rays. They do not necessarily have to result in a change of $f_{\rm T}$.}. 

In conclusion, we expect Eq. \ref{eq: trapped fraction fit} to be applicable for internal magnetic fields with $|B_r| \gtrsim |B_\theta|, |B_\phi|$ and a dipole-like dependence on the latitude (i.e., $B_r \propto \cos \theta$). 
Furthermore, Eq. \ref{eq: trapped fraction fit} is also prone to the limitations discussed in Sect. \ref{sect: limitations}.
These include the use of the WKBJ approximation and the fact that we only considered internal magnetic fields with a dipole-like dependence on latitude.

\subsection{Evolution of the trapped fraction} \label{sect: evolution f_T RGB}

\begin{figure*}[]
    \sidecaption
    \resizebox{\hsize}{!}{\includegraphics{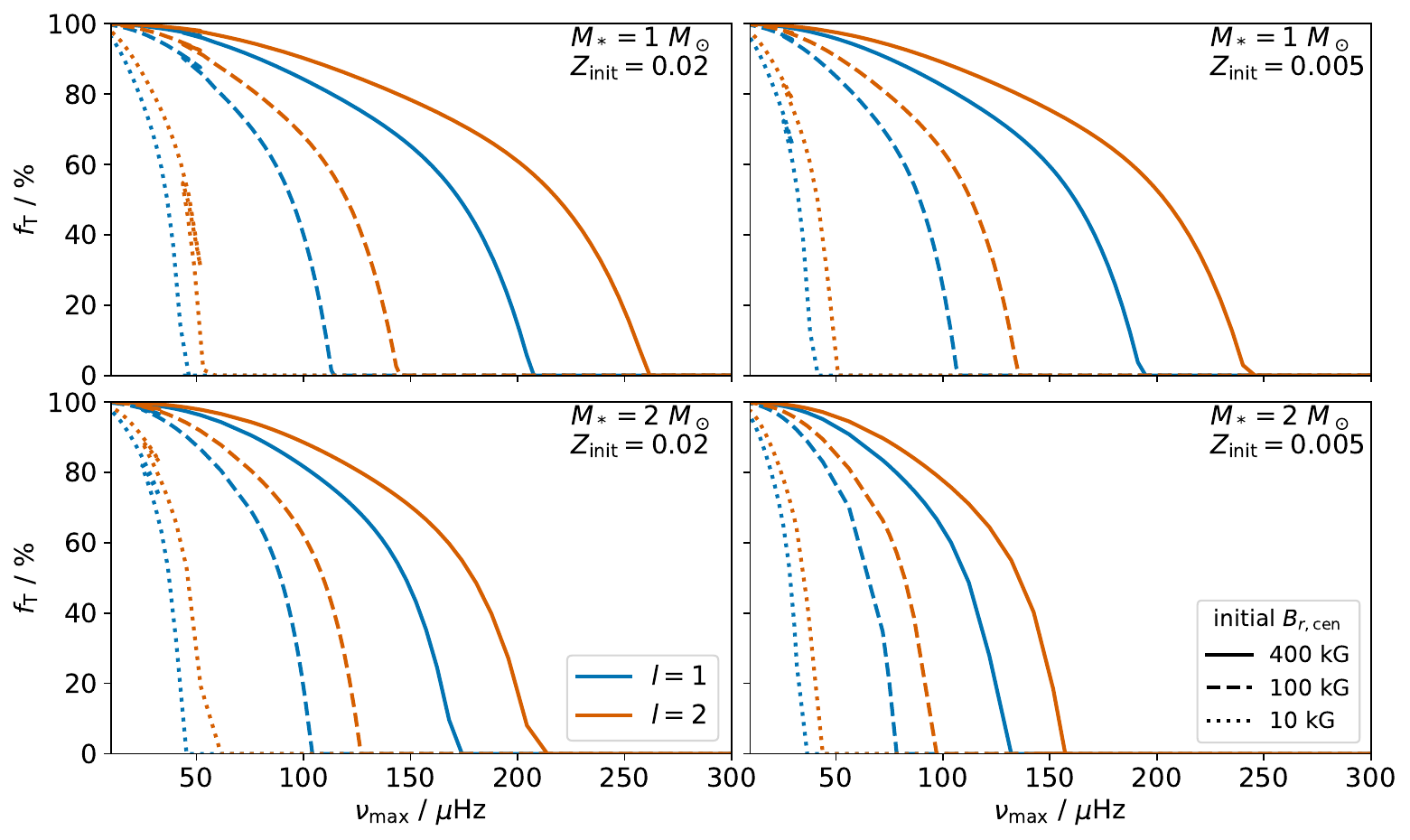}}
    \caption{Trapped fraction evaluated with Eq. \ref{eq: trapped fraction fit} at $\omega = 2\pi\nu_{\rm max}$, where $\nu_{\rm max}$ is the frequency of maximum oscillation power, as a function of the evolution of the star indicated by $\nu_{\rm max}$ (stars evolve from right to left). Each panel features one of the four evolutionary tracks presented in Fig. \ref{fig: HRD}. \textit{Rows} show tracks with the same stellar mass and \textit{columns} show tracks with the same metallicity. The blue (red) lines correspond to the dipole (quadrupole) modes. Different linestyles (solid, dashed, dotted) represent different initial strengths of the radial component of the magnetic field at the stellar center, which corresponds to the field strength the end of the MS. Due to the conservation of magnetic flux, the magnetic field strength increases as the stars evolve. The small knees that occur at the points where $f_{\rm T}$ becomes greater than 0, which are visible in some of the curves, are caused by the limited temporal resolution of the evolutionary tracks. We show the evolution of the trapped fraction considering $B_{\rm H} / B_{\rm crit,H}$ instead of $B_{\rm cen} / B_{\rm crit,cen}$ in Fig. \ref{fig: trapped fraction evol Hshell}.}
    \label{fig: trapped fraction evol center}
\end{figure*}

The critical field strength generally decreases as the star ascends the RGB (see Table \ref{tab: MESA models}). In addition, the strength of an internal magnetic field contained in the radiative core of an RGB star increases as the star evolves due to the conservation of magnetic flux and the contraction of the core region. Both of these effects cause the ratio $B_{\rm cen} / B_{\rm crit,cen}$ to grow, which means that $f_{\rm T}$ is expected to become higher during evolution in the RGB. In Fig. \ref{fig: trapped fraction evol center}, we show the evolution of the trapped fraction of both the dipole and quadrupole modes at $\omega = 2\pi\nu_{\rm max}$ for the four evolutionary tracks presented in Fig. \ref{fig: HRD}. For each track, we show the trapped fraction for three different central initial strengths of the magnetic field, which corresponds to the field strength at the end of the MS. These initial field strengths are scaled according to the conservation of magnetic flux during the evolution of the star.

As expected, $f_{\rm T}$ is initially 0 \% for all of our evolutionary tracks because the ratio $B_{\rm cen} / B_{\rm crit,cen}$ is at first small. If the magnetic field strength has become strong enough, $f_{\rm T}$ increases and approaches 100 \%. The onset of the trapping occurs earlier in the evolution of the star when the initial field strength is higher. Similarly, the suppression of the quadrupole mode always starts at higher values of the frequency of maximum oscillation power $\nu_{\rm max}$ compared to the dipole modes when considering the same initial magnetic field strength. This is a direct consequence of Eq. \ref{eq: critical field strength}. 

The stellar mass also affects the trapped fraction, with trapping starting at higher values of $\nu_{\rm max}$ for lower mass stars. This effect becomes weaker as the star evolves. 
It is questionable whether this feature is observable, as the typical magnetic field strength at the end of the MS of stars with $M_* = 1\ M_\odot$ and $2\ M_\odot$ may be different.
In fact, the study by \citet{bugnet+21} suggests that this may not be possible. They estimate the strength of the fossil magnetic field left by a potential convection-driven dynamo phase and show that a change in stellar mass and metallicity does not significantly alter the magnetic frequency shift of mixed modes of similar frequency on the RGB. This implies that the effect of the internal magnetic field may be comparable for different stellar masses and metallicities at a certain $\nu_{\rm max}$\footnote{Note that the analysis of \citet{bugnet+21} is based on a first-order perturbation approach, which is technically not applicable to the magnetic field strengths discussed in the present work.}. 
The influence of stellar metallicity on the value of $f_{\rm T}$ in Fig. \ref{eq: critical field strength} appears to be smaller than that of the initial field strength, the spherical degree, and the stellar mass.

The values selected for the initial magnetic field strength range from 10 to 400 kG. They are an order of magnitude larger than the MS field strengths inferred from the magnetic frequency shifts of the dipole modes by \citet{deheuvels+23} for an axisymmetric dipolar field configuration. However, these field strengths cannot be directly compared to each other. This is because the central field strength of the Prendergast field is the maximum strength of the magnetic field, while the field strengths estimated by \citet{deheuvels+23} are weighted averages dominated by the contribution of the hydrogen-burning shell \citep[e.g.,][]{li+22, bhattacharya+24}.
In particular, the radial component of the magnetic field in the case of the Prendergast field decreases rapidly with increasing distance from the center, while a uniform distribution is assumed for the estimation of the MS field strengths by \citet{deheuvels+23}. Therefore, the central field strengths of the Prendergast field required to suppress the multipole modes are higher than the average values estimated from the observed frequency shifts.

The improvised magnetic field introduced in Sect. \ref{sect: applicability f_T} is more similar to the field configuration assumed by \citet{deheuvels+23}, because the radial component of the field is uniform in the region at and interior to the hydrogen-burning shell. Since the improvised field is scaled by the field strength at the hydrogen-burning shell $B_{\rm H}$, the MS field strengths estimated by \citet{deheuvels+23} should be compared with $B_{\rm crit,H}$ instead of $B_{\rm crit,cen}$. 
In Fig. \ref{fig: trapped fraction evol Hshell}, we show that the evolution of the trapped fraction is comparable to that in Fig. \ref{fig: trapped fraction evol center} when using initial field strengths ranging from 1 to 40 kG and the ratio $B_{\rm H} / B_{\rm crit,H}$ instead of $B_{\rm cen} / B_{\rm crit,cen}$. This indicates that the MS field strengths estimated by \citet{deheuvels+23} might be sufficient to cause dipole mode suppression starting from $\nu_{\rm max} \approx 250\ \mu$Hz, which is in good agreement with observations \citep[e.g.,][]{stello+16b, mosser+17, coppee+24}. Outliers from the population of suppressed dipole mode stars with higher values of $\nu_{\rm max}$ could be caused, for example, by slightly higher MS field strengths or different field configurations.

\section{Discussion} \label{sect: Discussion}

\subsection{Implications for observed power spectral densities} \label{sect: comparision observations}

\begin{figure}[]
    \centering
    \resizebox{\hsize}{!}{\includegraphics{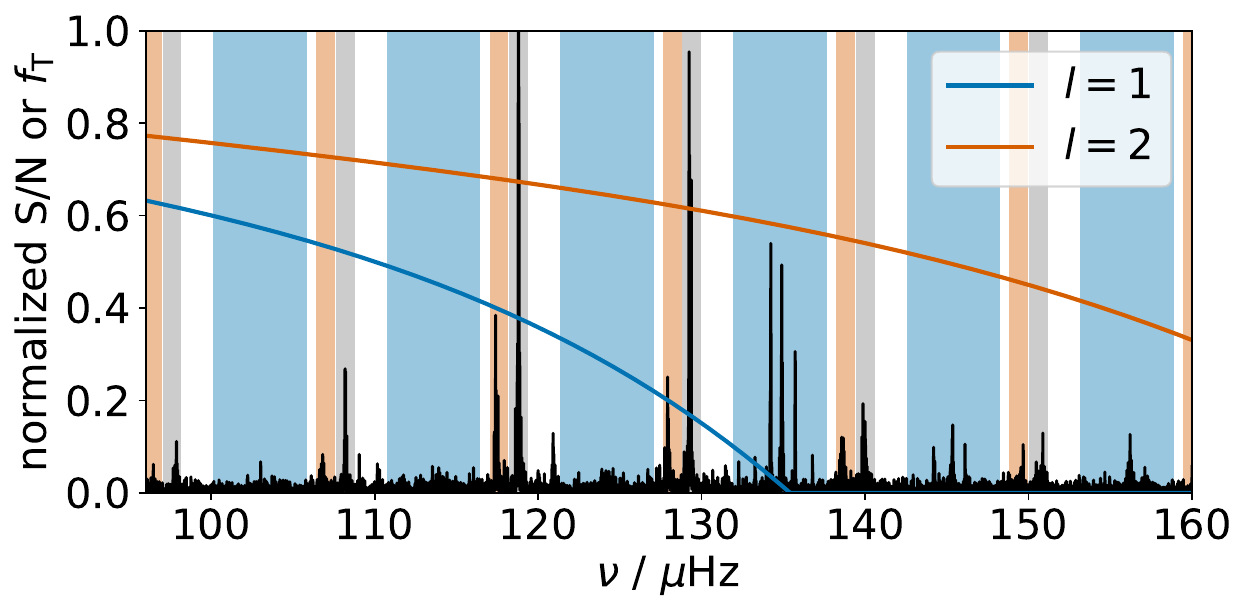}}
    \caption{Background-corrected PSD of KIC 6975038 obtained from {\it Kepler} data using the background model of \citet{kallinger+14} \citep[for details, see][]{coppee+24}. Radial modes are indicated by the gray shaded area, dipole modes by the blue area, and quadrupole modes by the reddish area. In addition, we show the trapped fraction as a function of frequency computed for a stellar model with $M_* = 1.3\ M_\odot$, $Z_{\rm init} = 0.02$, $\nu_{\rm max} \approx 128$ $\mu$Hz, and $B_{\rm cen} \approx 6.23$ MG (see Sect. \ref{sect: kic 6975038}) for both the dipole modes (in blue) and quadrupole modes (in red).}
    \label{fig: PSD KIC 6975038}
\end{figure}

The trapped fraction obtained in this work is not a quantity that can be measured directly in observations. Instead, the suppressed dipole mode stars are classified based on the mode visibility, which is defined as the ratio of the amplitude of the multipole modes of a spherical degree $l$ to the radial mode of the same pressure radial order \citep[e.g.,][]{fuller+15, stello+16b, cantiello+16}. The mode visibility depends on the trapped fraction, but there is currently no general expression linking these two quantities. As of yet, this is only possible for the two boundary cases, $f_{\rm T} = 0\ \%$ and $f_{\rm T} = 100\ \%$ \citep[e.g.,][]{fuller+15, takata16b, mosser+17}. If $f_{\rm T} = 0\ \%$, the mixed modes are not suppressed and have normal amplitudes. If, on the other hand, $f_{\rm T} = 100\ \%$, all the energy that leaks into the gravity mode cavity is dissipated by the internal magnetic field, which corresponds to the case discussed by \citet{fuller+15} \citep[see also][]{shibahashi79, unno+89}. For intermediate values of $f_{\rm T}$, the observed visibility is expected to lie between these two limiting cases.
Therefore, we can make qualitative comparisons between the expected behavior of $f_{\rm T}$ and the observed mode visibilities.

The results of our magneto-gravity ray tracing analysis show that the trapped fraction is a function of the frequency $\nu = \omega/(2\pi)$. 
For a stellar model with an internal magnetic field of a certain strength, $f_{\rm T}$ is equal to 0 \% at higher frequencies and increases at frequencies below a threshold value. It is therefore expected that the suppression of the multipole modes occurs at lower frequencies. It depends on the respective model and magnetic field strength whether the suppression falls within the observable frequency range.
The PSD of the RGB star KIC 6975038 (Fig. \ref{fig: PSD KIC 6975038}) features clear signs of dipole mode suppression at lower values of $\nu$, while the amplitudes of the dipole modes appear to be normal at higher $\nu$.
This phenomenon has been associated with the onset of the dipole mode suppression at lower $\nu$ caused by an internal magnetic field \citep[e.g.,][]{mosser+17, loi20a, deheuvels+23}. The case of KIC 6975038 is discussed in more detail in Sect. \ref{sect: kic 6975038}.

As soon as $f_{\rm T}$ is greater than 0 \%, it initially increases rapidly while $\nu$ decreases. Eventually, this sharp increase flattens out when $f_{\rm T}$ begins to approach 100 \%. 
The flattening of the frequency dependence of the trapped fraction is consistent with observations. So far, no trend in the observed visibilities with the frequency has been detected for the suppressed dipole mode stars, unless they exhibit features expected to be associated with the onset of dipole mode suppression (such as KIC 6975038).

The trapped fraction also depends on the spherical degree $l$. Our computations show that modes of higher $l$ are suppressed at lower magnetic field strengths. For the PSD of a star, this means that the suppression of quadrupole modes starts at a higher frequency than the suppression of dipole modes \citep[see Fig. \ref{fig: PSD KIC 6975038} and][]{loi20a}.
In observations, however, the visibility of the quadrupole modes of the suppressed dipole mode stars is usually higher than that of the dipole modes. This is because the mode visibility depends not only on the trapped fraction, but also on the coupling strength of the mixed modes \citep[e.g.,][]{shibahashi79, fuller+15, takata16a, takata16b, mosser+17}. The coupling strength is a measure of how strongly the oscillations in the pressure and gravity mode cavities interact. At higher spherical degrees, the interaction between the mode cavities is weaker \citep[e.g.,][]{hekker+17}, such that the coupling strength is lower and the mode visibility is higher \citep[e.g.,][]{cantiello+16}.
Overall, our results suggest that the quadrupole modes of stars with suppressed dipole modes should also be suppressed, although their mode visibilities might be higher.

\subsection{Trapped fraction along the red-giant branch}  \label{sect: evolution f_T discussion}

The observed dipole mode visibilities of RGB stars show a clear dichotomy. On one hand, there is a population of stars with normal dipole mode visibilities. One the other hand, there are the suppressed dipole mode stars, which have been observed for $\nu_{\rm max} \lesssim 250\ \mu$Hz \citep{stello+16b, mosser+17}. While the two populations start to merge at $\nu_{\rm max} \lesssim 70\ \mu$Hz \citep{stello+16b}, there are only a few stars with $\nu_{\rm max} > 70\ \mu$Hz whose dipole mode visibility lies between the two branches and cannot be assigned to either of them.

In Sect. \ref{sect: evolution f_T RGB}, we investigated the evolution of the trapped fraction under the assumption of conservation of magnetic flux for different magnetic field strengths at the end of the MS. We have shown that for stars with solar mass, the dipole mode suppression can start at $\nu_{\rm max} \lesssim 200\ \mu$Hz if the central field strength at the end of the MS is $\lesssim 400$ kG. If the strength of the field at the hydrogen-burning shell is considered instead, a field strength of $\lesssim 40$ kG at the end of the MS is sufficient to cause dipole mode suppression at $\nu_{\rm max} \lesssim 250\ \mu$Hz. 
These magnetic field strengths at the hydrogen-burning shell agree with those inferred by \citet{deheuvels+23} from observed average frequency shifts of the dipole modes of RGB stars assuming an axisymmetric dipolar magnetic field configuration. Note however that these field strengths might be overestimated \citep{rui+23}.

The small number of stars with observed dipole mode visibilities that lie between the visibilities of the population of normal RGB stars and the population of suppressed dipole mode stars could be related to the sharp increase in $f_{\rm T}$ immediately after the onset of dipole mode suppression (see Figs. \ref{fig: trapped fraction evol center} and \ref{fig: trapped fraction evol Hshell}). It is not unreasonable to assume that within the uncertainties, there is a threshold value of $f_{\rm T}$ above which the dipole mode visibility is indistinguishable from that of a star experiencing complete dipole mode suppression (i.e., $f_{\rm T} = 100\ \%$).
Once $f_{\rm T} > 0 \%$, the trapped fraction could quickly reach this threshold as the star ascends the RGB since $f_{\rm T}$ rises sharply after the start of the dipole mode suppression. 
This would lead to a short time window for the detection of an RGB star with an intermediate visibility in terms of stellar evolution, which is in line with the current state of observations.

Contrary to \citet{loi20a}, we find that the trapped fraction always follows Eq. \ref{eq: trapped fraction fit}, even for more evolved RGB stars. However, the observed dipole mode visibilities of the suppressed dipole mode stars increase with decreasing $\nu_{\rm max}$ \citep{stello+16b, mosser+17}. This trend can simply be explained by the decrease in coupling strength during the evolution of an RBG star \citep{cantiello+16}. The fact that $f_{\rm T}$ always saturates at 100 \% in our simulations therefore does not contradict the observed behavior of dipole mode visibility.

Lastly, assuming that all RGB stars have an internal magnetic field of a certain strength at the end of the MS, we expect the fraction of stars with suppressed dipole modes to increase with decreasing $\nu_{\rm max}$. This is because the trapped fraction increases as the star evolves along the RGB (see Figs. \ref{fig: trapped fraction evol center} and \ref{fig: trapped fraction evol Hshell}). However, this trend could be obscured in the observations due to the generally higher dipole mode visibility of the suppressed dipole mode stars at low $\nu_{\rm max}$ \citep[see][]{stello+16b, mosser+17}, which is caused by the decreasing coupling strength. Furthermore, the radiative damping in the core of RGB stars becomes stronger at lower $\nu_{\rm max}$ \citep[e.g.,][Bordadágua et al. in prep.]{dupret+09, grosjean+14, hekker+17}, which makes the identification of effects caused by an internal magnetic field more challenging.

\subsection{Frequency shift of KIC 6975038} \label{sect: kic 6975038}

\begin{table*}[]
    \caption{Average frequency shifts of the dipole modes caused by different internal magnetic field configurations for three stellar models.}
    \centering
    \begin{tabular}{c|c|c|c|c|c|c|c}
    \hline\hline
    $M_*\ /\ M_\odot$ & $Z_{\rm init}$ & $\nu_{\rm max}\ /\ \mu$Hz & $B_{\rm crit,cen}\ /\ $MG & $B_{\rm crit,H}\ /\ $MG & $\left(\frac{\delta\omega_0}{2\pi}\right)_{\rm PG,H}\ /\ \mu$Hz & $\left(\frac{\delta\omega_0}{2\pi}\right)_{\rm PG,core}\ /\ \mu$Hz & $\left(\frac{\delta\omega_0}{2\pi}\right)_{\rm pla}\ /\ \mu$Hz \\
    \hline
    1 & 0.02 & 127.03 & 8.78 & 0.39 & 1.33 (2.65) & 1.40 (2.80) & 2.09 (4.18) \\
    1 & 0.005 & 129.53 & 8.87 & 0.47 & 1.36 (2.72) & 1.40 (2.81) & 2.38 (4.75) \\
    1.3 & 0.02 & 127.99 & 8.90 & 0.35 & 1.34 (2.69) & 1.36 (2.71) & 2.06 (4.13) \\
    \hline\hline
    \end{tabular}
    \tablefoot{The models were selected from each evolutionary track to mimic the observed value of $\nu_{\rm max}$ of KIC 6975038 \citep[e.g.,][]{deheuvels+23}. The magnetic field strengths have been chosen such that the suppression of the multipole modes behaves as shown in Fig. \ref{fig: PSD KIC 6975038}. The average frequency shifts have been determined using the first-order perturbation approach of \citet{li+22}. Note that the application of this method to magnetic field strengths close to the critical field strength is expected to lead to an underestimation of the frequency shifts \citep{rui+23}. The quantity $(\delta\omega_0 /2\pi)_{\rm PG,H}$ corresponds to the frequency shift induced by a Prendergast field that is contained by the region at and within the hydrogen-burning shell, while $(\delta\omega_0 /2\pi)_{\rm PG,core}$ corresponds to a Prendergast field that extends over the entire radiative core. $(\delta\omega_0 /2\pi)_{\rm pla}$ corresponds to the frequency shift induced by the improvised field with a plateau within the hydrogen-burning shell introduced in Sect. \ref{sect: applicability f_T}.
    The value of the average frequency shifts without brackets (in brackets) corresponds to the azimuthal order $m=0$ ($m = \pm 1$).}
    \label{tab: freq shift kic 6975038}
\end{table*}

\begin{figure}[]
    \centering
    \resizebox{\hsize}{!}{\includegraphics{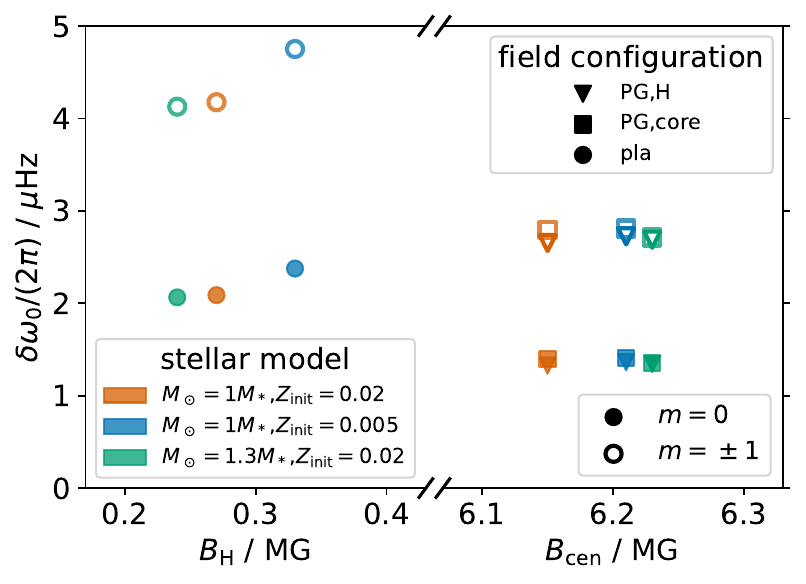}}
    \caption{Average frequency shift of the dipole modes caused by various internal magnetic field configurations as a function of the magnetic field strength for three stellar models. The field strengths have been chosen such that $B_{\rm cen} = 0.7\ B_{\rm crit,cen}$ and $B_{\rm H} = 0.7\ B_{\rm crit,H}$. The critical field strengths were calculated at $\omega = 2\pi\nu_{\rm max}$ with $l=1$. Colors correspond to different stellar models and symbols indicate that a different magnetic field configuration has been used. Average frequency shifts corresponding to the azimuthal order $m=0$ ($m = \pm1$) are shown as solid symbols (outlines). The field configurations and the calculation of the average frequency shifts are described in Sect. \ref{sect: kic 6975038} and in the notes under Table \ref{tab: freq shift kic 6975038}, which also summarizes the data shown in this figure. Note that the horizontal axis is broken.}
    \label{fig: frequency shift KIC 6975038}
\end{figure}

The star KIC 6975038 is of particular interest for the hypothesis of an internal magnetic field, because the dipole modes are suppressed at lower frequencies, while their amplitudes appear to be normal at higher frequencies (see Fig. \ref{fig: PSD KIC 6975038}). \citet{deheuvels+23} show that the dipole modes split into asymmetric multiplets at higher frequencies, of which one component per multiplet was identified. They measure the average frequency shift of the dipole modes at $\nu_{\rm max}$ as $\delta\omega_0/(2\pi) = 7.18 \pm 0.03$ $\mu$Hz and attribute it to the presence of an internal magnetic field. Since only one component of each multiplet has been identified, the configuration of the internal magnetic field of KIC 6975038 is currently not constrained \citep{deheuvels+23}.

Here, we used the first-order perturbation approach presented by \citet{li+22} to calculate the expected average frequency shift for a selected stellar model with an internal magnetic field \citep[see also][]{bugnet+21, loi+21, bugnet22}. The average frequency shift only depends on the radial component of the magnetic field. 
It is important to note that this approach assumes that the magnetic field only constitutes a small perturbation to the restoring forces of the oscillations. This assumption is broken when the field strength approaches the critical field strength.
Therefore, our predictions of the average frequency shift are expected be underestimated \citep{rui+23}. 
This means that our estimates of the observed average frequency shift must be lower than the observed value, but likely of the same order of magnitude \citep{rui+23}, to be consistent with the current understanding of magnetic frequency shifts.

We set the internal field strength to 70 \% of the critical field strength of the dipole modes at $\omega = 2\pi\nu_{\rm max}$, so that the onset of dipole mode suppression is at about $\nu \approx 135\ \mu$Hz (see Fig. \ref{fig: PSD KIC 6975038}). 
\citet{yu+18} estimate the mass of KIC 6975038 to be around $1.3\ M_\odot$. Therefore, we computed an additional evolutionary track with this mass and $Z_{\rm init} = 0.02$ using the MESA inlist shown in Appendix \ref{app: MESA inlist}. 
We then selected a model with $\nu_{\rm max} \approx 128$ $\mu$Hz from this track, which is close to the observed value of $\nu_{\rm max}$ \citep[e.g.,][]{deheuvels+23}. In Fig. \ref{fig: PSD KIC 6975038}, we show $f_{\rm T}$ as a function of $\nu$ for this model.

We tested three topologies of the internal magnetic field: two Prendergast configurations and the improvised field introduced in Sect. \ref{sect: applicability f_T}. They were scaled such that $B_{\rm cen} = 0.7\ B_{\rm crit,cen}$ and $B_{\rm H}^{\rm pla} = 0.7\ B_{\rm crit,H}$, respectively.
When considering Prendergast fields, the average frequency shift depends on the maximum extent of the field $r_{\rm f}$. We investigated two scenarios: A Prendergast field that is confined by the region at and within the hydrogen-burning shell and a Prendergast field that extends over the entire radiative core. Since the magnetic frequency shift is only sensitive to the region at and within the hydrogen-burning shell \citep[e.g.,][]{li+22, bhattacharya+24}, the average frequency shift induced by the improvised field does not change significantly when the maximum extent of the field is changed, as it behaves like a plateau in this region.

In Table \ref{tab: freq shift kic 6975038} and Fig. \ref{fig: frequency shift KIC 6975038}, we show magnetic field strengths and the resulting average frequency shifts of the dipole modes for the three different field configurations for the azimuthal orders $m = 0$ and $m = \pm 1$ \citep[see][]{li+22, deheuvels+23}\footnote{The asymmetry parameter is $a = 2/5$ for all tested magnetic field configurations.}.
Note that due to the limitations of the Hamiltonian ray tracing approach described in Sect. \ref{sect: limitations} and the simplicity of the stellar model, the magnetic field strengths determined for KIC 6975038 should not be considered as measurements, but as an order of magnitude estimate (see Sect. \ref{sect: limitations}).
Since recent studies estimate the mass of KIC 6975038 to be less than $1.3\ M_\odot$ (e.g., Coppée et al. in prep.), we repeated the process described above for a model of the evolutionary track with $M_* = 1\ M_\odot$ and $Z_{\rm init} = 0.02$, as well as for a model of the one with $M_* = 1\ M_\odot$ and $Z_{\rm init} = 0.005$. The field strengths and average frequency shifts inferred for these models are also shown in Table \ref{tab: freq shift kic 6975038} and Fig. \ref{fig: frequency shift KIC 6975038}.

All of our estimates for the average frequency shift are systematically lower than the observed value. However, they agree with it by less than a factor of 5.5.
Taking into account the limitations of the first-order perturbation approach and the uncertainties in the estimation of the magnetic field strength based on the ray tracing method, all tested topologies of the internal magnetic field are in agreement with the observations. Based on these results, we cannot favor any of the field configurations over the others. 
Overall, our order-of-magnitude analysis suggests that both the suppression of the dipole modes at lower $\nu$ and the frequency shift of the modes at higher $\nu$ can be explained by the presence of an internal magnetic field. 
Our results thus support the hypothesis that the cause of these phenomena is indeed magnetic in nature.
However, a more sophisticated treatment of the magnetic effects is required to firmly validate this interpretation of the features observed in the oscillations of KIC 6975038.

\subsection{Limitations} \label{sect: limitations}
   
As mentioned in Sect. \ref{sect: hamilton}, the Hamiltonian ray tracing approach used to estimate the trapped fractions relies on the WKBJ approximation of zeroth order. This means that higher order effects, such as the evanescence of the oscillation in certain regions of the star, mode excitation and damping, as well as phase dependencies, are not taken into account. However, none of these effects should have an influence on the trapped fraction \citep[see also][]{loi20a}.
More important for our application is the assumption that the horizontal wavelength is much smaller than the horizontal variation of the internal magnetic field. Since the observable oscillation modes have a low spherical degree, this assumption is violated in our procedure and could lead to a bias in our estimates of the trapped fraction.

Nevertheless, there are a number of arguments that suggest that Eq. \ref{eq: trapped fraction fit} describes the trapped fraction with reasonable accuracy. 
First, the trapped fraction is an increasing function of the strength of the internal magnetic field and transitions smoothly from 0 \% to 100 \%. This is exactly what is needed to explain the observation that some of the suppressed dipole mode stars have dipole modes with a mixed character \citep[][Coppée et al. in prep.]{mosser+17}. 
Second, the transition from 0 \% to 100 \% takes place when the strength of the internal magnetic field approaches the critical field strength. This was predicted by \citet{fuller+15} and later verified by \citet{rui+fuller23}. 
Third, the average frequency shift estimated for KIC 6975038 corresponding to the field strengths obtained with Eq. \ref{eq: trapped fraction fit} is systematically lower than the average frequency shift observed by \citet{deheuvels+23}. This is in line with the results of \citet{rui+23}.
Fourth, Eq. \ref{eq: trapped fraction fit} is a good fit for the trapped fraction as a function of the internal field strength also for rays with a high spherical degree of $l = 10$ and $l = 100$ (see Fig. \ref{fig: trapped fraction l=10 l=100}). These rays have a much smaller horizontal wavelength and therefore correspond more to the short wavelength assumption. By extrapolation, we conclude that Eq. \ref{eq: trapped fraction fit} should also apply to the observable modes with smaller $l$, since it appears to be independent of the spherical degree.
However, in order to verify these results, more sophisticated simulations of the interaction of the waves with the internal magnetic field are required.  As long as this is not the case, Eq. \ref{eq: trapped fraction fit} should be considered as an order of magnitude estimate that can be used to describe general qualitative trends of the multipole mode suppression.

A further limitation arises from the fact that we have only considered internal magnetic fields with a dipole-like dependence on the latitude. In reality, other field geometries are possible and have been observed \citep{li+22, li+23, hatt+24}.
The applicability of Eq. \ref{eq: trapped fraction fit} to different field configurations was discussed in Sect. \ref{sect: applicability f_T} and we refer the reader to this section for more information. In short, we expect a different behavior of the trapped fraction for magnetic field topologies with higher order contributions such as a quadrupole component. However, it can be assumed that the transition of the trapped fraction from 0 \% to 100 \% takes place when the internal field strength approaches the critical field strength, which means that the resulting changes are most likely of a quantitative nature.

Finally, we have neglected rotation. The typical rotation frequencies of RGB stars are small compared to the frequencies of the oscillations. Therefore, the inclusion of rotation is not expected to significantly affect our conclusions \citep[see also][]{loi20a, loi20b}, although this remains to be verified.
For the dipole modes, \citet{loi20b} speculate that the internal magnetic field could primarily damp the zonal oscillations (i.e., the component with $m = 0$ with respect to the magnetic axis). The reason for this is that the reflected rays are those that propagate in an equatorial band, which corresponds to sectoral oscillations.
In the particular case where the magnetic axis and the rotation axis are misaligned, the oscillation modes are expected to separate into $(2 l +1)^2$ individual peaks \citep[e.g.,][]{gough&thompson90, loi+21}. However, following the argument of \citet{loi20b}, only the rotationally split peak corresponding to $m=\pm1$ with respect to the magnetic axis could be visible. This would reduce the number of observable modes.

\section{Summary}  \label{sect: Summary}

In this work, we built on the study of \citet{loi20a} and performed a Hamiltonian ray tracing analysis of magneto-gravity waves using numerical models of RGB stars with different masses, metallicities and ages as background.
From these simulations, we estimated the trapped fraction, which is a measure of the mode energy lost through interaction with an internal magnetic field.
We confirm the results of \citet{loi20a} that the trapped fraction transitions from 0 \% to 100 \% with increasing strength of the internal magnetic field. In particular, the trapped fraction depends on the magnetic field strength, the frequency of the oscillations, the spherical degree, and the background model (Sect. \ref{sect: results ray tracing}). However, we do not reproduce the result of \citet{loi20a} that the trapped fraction saturates at values smaller than 100 \% for more evolved RGB stars. This could be a consequence of the changes we have made to the integration of the rays (Sect. \ref{sect: ray initialization}) and the ray classification scheme (Appendix \ref{app: ray classification}).

Furthermore, we find that these dependencies are encompassed by the dependence of the critical magnetic field strength (Eq. \ref{eq: critical field strength}) on these parameters.
We show that the trapped fraction depends only on the ratio of the central strength of the internal magnetic field to the central critical field strength. The behavior of the trapped fraction is described by Eq. \ref{eq: trapped fraction fit} for dipole-like configurations of the internal magnetic field.
In addition, we show in Sect. \ref{sect: applicability f_T} that Eq. \ref{eq: trapped fraction fit} is also applicable when the ratio of the magnetic field strength to the critical field strength is evaluated at the location of the hydrogen-burning shell where the critical field strength has a minimum, instead of at the center. Our results suggest that the ratio of the internal field strength to the critical field strength at the location where the field strength first approaches the critical field strength determines the trapped fraction.

Equation \ref{eq: trapped fraction fit} provides a flexible tool that can be used to improve our understanding of the multipole mode suppression observed for RGB stars. It can, for example, be used to describe the trapped fraction as a star ascends the RGB, taking into account the evolution of the strength of the internal magnetic field (Sect. \ref{sect: evolution f_T RGB}).
In addition, Eq. \ref{eq: trapped fraction fit} can be used to study the onset of mode suppression in the observed power spectra densities of RGB stars (Sect. \ref{sect: kic 6975038}).
Note however that due to the limitations of the Hamiltonian ray tracing approach, Eq. \ref{eq: trapped fraction fit} should be regarded as an order of magnitude estimate and needs to be validated by future studies using a more complete treatment of the magnetic effects (for details, see Sect. \ref{sect: limitations}).

Based on our results and those of previous studies \citep{fuller+15, mosser+17, loi20a}, we predict that the quadrupole modes of stars with suppressed dipole modes should always be suppressed (Sect. \ref{sect: comparision observations}). In addition, we expect that the ratio of suppressed dipole modes to RGB stars with normal dipole mode amplitudes should increase along the RGB (i.e., with decreasing $\nu_{\rm max}$). However, this trend could be obscured by the generally higher dipole mode visibilities of the suppressed dipole mode stars at low $\nu_{\rm max}$ \citep[Sect. \ref{sect: evolution f_T discussion};][]{stello+16b, cantiello+16}.

In the next article in this series, we will investigate the relationship between intermediate values of the trapped fraction (i.e., $0\ \% < f_{\rm T} < 100\ \%$) and the observable visibility of the multipole modes. Furthermore, we will relate the different morphologies observed in the PSD of stars with suppressed dipole modes (Coppée et al. in prep.) to physical quantities.

\begin{acknowledgements}
We thank the anonymous referee for their valuable and constructive feedback that improved the manuscript. We acknowledge funding from the ERC Consolidator Grant DipolarSound (grant agreement \# 101000296). In addition, we acknowledge support from the Klaus Tschira Foundation. 
\end{acknowledgements}

\bibliographystyle{aa}
\bibliography{ref}

\begin{thebibliography}{72}
\expandafter\ifx\csname natexlab\endcsname\relax\def\natexlab#1{#1}\fi

\bibitem[{{Akg{\"u}n} {et~al.}(2013){Akg{\"u}n}, {Reisenegger}, {Mastrano}, \& {Marchant}}]{akgun+13}
{Akg{\"u}n}, T., {Reisenegger}, A., {Mastrano}, A., \& {Marchant}, P. 2013, \mnras, 433, 2445

\bibitem[{{Becerra} {et~al.}(2022){Becerra}, {Reisenegger}, {Valdivia}, \& {Gusakov}}]{becerra+22}
{Becerra}, L., {Reisenegger}, A., {Valdivia}, J.~A., \& {Gusakov}, M. 2022, \mnras, 517, 560

\bibitem[{{Beck} {et~al.}(2014){Beck}, {Hambleton}, {Vos}, {Kallinger}, {Bloemen}, {Tkachenko}, {Garc{\'\i}a}, {{\O}stensen}, {Aerts}, {Kurtz}, {De Ridder}, {Hekker}, {Pavlovski}, {Mathur}, {De Smedt}, {Derekas}, {Corsaro}, {Mosser}, {Van Winckel}, {Huber}, {Degroote}, {Davies}, {Pr{\v{s}}a}, {Debosscher}, {Elsworth}, {Nemeth}, {Siess}, {Schmid}, {P{\'a}pics}, {de Vries}, {van Marle}, {Marcos-Arenal}, \& {Lobel}}]{beck+14}
{Beck}, P.~G., {Hambleton}, K., {Vos}, J., {et~al.} 2014, \aap, 564, A36

\bibitem[{{Beck} {et~al.}(2012){Beck}, {Montalban}, {Kallinger}, {De Ridder}, {Aerts}, {Garc{\'\i}a}, {Hekker}, {Dupret}, {Mosser}, {Eggenberger}, {Stello}, {Elsworth}, {Frandsen}, {Carrier}, {Hillen}, {Gruberbauer}, {Christensen-Dalsgaard}, {Miglio}, {Valentini}, {Bedding}, {Kjeldsen}, {Girouard}, {Hall}, \& {Ibrahim}}]{beck+2012}
{Beck}, P.~G., {Montalban}, J., {Kallinger}, T., {et~al.} 2012, \nat, 481, 55

\bibitem[{{Bedding} {et~al.}(2011){Bedding}, {Mosser}, {Huber}, {Montalb{\'a}n}, {Beck}, {Christensen-Dalsgaard}, {Elsworth}, {Garc{\'\i}a}, {Miglio}, {Stello}, {White}, {De Ridder}, {Hekker}, {Aerts}, {Barban}, {Belkacem}, {Broomhall}, {Brown}, {Buzasi}, {Carrier}, {Chaplin}, {di Mauro}, {Dupret}, {Frandsen}, {Gilliland}, {Goupil}, {Jenkins}, {Kallinger}, {Kawaler}, {Kjeldsen}, {Mathur}, {Noels}, {Silva Aguirre}, \& {Ventura}}]{bedding+11}
{Bedding}, T.~R., {Mosser}, B., {Huber}, D., {et~al.} 2011, \nat, 471, 608

\bibitem[{{Bhattacharya} {et~al.}(2024){Bhattacharya}, {Das}, {Bugnet}, {Panda}, \& {Hanasoge}}]{bhattacharya+24}
{Bhattacharya}, S., {Das}, S.~B., {Bugnet}, L., {Panda}, S., \& {Hanasoge}, S.~M. 2024, \apj, 970, 42

\bibitem[{{Borucki} {et~al.}(2010){Borucki}, {Koch}, {Basri}, {Batalha}, {Brown}, {Caldwell}, {Caldwell}, {Christensen-Dalsgaard}, {Cochran}, {DeVore}, {Dunham}, {Dupree}, {Gautier}, {Geary}, {Gilliland}, {Gould}, {Howell}, {Jenkins}, {Kondo}, {Latham}, {Marcy}, {Meibom}, {Kjeldsen}, {Lissauer}, {Monet}, {Morrison}, {Sasselov}, {Tarter}, {Boss}, {Brownlee}, {Owen}, {Buzasi}, {Charbonneau}, {Doyle}, {Fortney}, {Ford}, {Holman}, {Seager}, {Steffen}, {Welsh}, {Rowe}, {Anderson}, {Buchhave}, {Ciardi}, {Walkowicz}, {Sherry}, {Horch}, {Isaacson}, {Everett}, {Fischer}, {Torres}, {Johnson}, {Endl}, {MacQueen}, {Bryson}, {Dotson}, {Haas}, {Kolodziejczak}, {Van Cleve}, {Chandrasekaran}, {Twicken}, {Quintana}, {Clarke}, {Allen}, {Li}, {Wu}, {Tenenbaum}, {Verner}, {Bruhweiler}, {Barnes}, \& {Prsa}}]{borucki+10}
{Borucki}, W.~J., {Koch}, D., {Basri}, G., {et~al.} 2010, Science, 327, 977

\bibitem[{{Braithwaite}(2006)}]{braithwaite06}
{Braithwaite}, J. 2006, \aap, 453, 687

\bibitem[{{Braithwaite}(2007)}]{braithwaite07}
{Braithwaite}, J. 2007, \aap, 469, 275

\bibitem[{{Braithwaite}(2008)}]{braithwaite08}
{Braithwaite}, J. 2008, \mnras, 386, 1947

\bibitem[{{Braithwaite}(2009)}]{braithwaite09}
{Braithwaite}, J. 2009, \mnras, 397, 763

\bibitem[{{Braithwaite} \& {Nordlund}(2006)}]{braithwaite+nordlund06}
{Braithwaite}, J. \& {Nordlund}, {\r{A}}. 2006, \aap, 450, 1077

\bibitem[{{Braithwaite} \& {Spruit}(2017)}]{braithwaite+spruit17}
{Braithwaite}, J. \& {Spruit}, H.~C. 2017, Royal Society Open Science, 4, 160271

\bibitem[{{Bugnet}(2022)}]{bugnet22}
{Bugnet}, L. 2022, \aap, 667, A68

\bibitem[{{Bugnet} {et~al.}(2021){Bugnet}, {Prat}, {Mathis}, {Astoul}, {Augustson}, {Garc{\'\i}a}, {Mathur}, {Amard}, \& {Neiner}}]{bugnet+21}
{Bugnet}, L., {Prat}, V., {Mathis}, S., {et~al.} 2021, \aap, 650, A53

\bibitem[{{Cantiello} {et~al.}(2016){Cantiello}, {Fuller}, \& {Bildsten}}]{cantiello+16}
{Cantiello}, M., {Fuller}, J., \& {Bildsten}, L. 2016, \apj, 824, 14

\bibitem[{{Copp{\'e}e} {et~al.}(2024){Copp{\'e}e}, {M{\"u}ller}, {Bazot}, \& {Hekker}}]{coppee+24}
{Copp{\'e}e}, Q., {M{\"u}ller}, J., {Bazot}, M., \& {Hekker}, S. 2024, \aap, 690, A324

\bibitem[{{Das} {et~al.}(2024){Das}, {Einramhof}, \& {Bugnet}}]{das+24}
{Das}, S.~B., {Einramhof}, L., \& {Bugnet}, L. 2024, \aap, 690, A217

\bibitem[{{Deheuvels} {et~al.}(2014){Deheuvels}, {Do{\u{g}}an}, {Goupil}, {Appourchaux}, {Benomar}, {Bruntt}, {Campante}, {Casagrande}, {Ceillier}, {Davies}, {De Cat}, {Fu}, {Garc{\'\i}a}, {Lobel}, {Mosser}, {Reese}, {Regulo}, {Schou}, {Stahn}, {Thygesen}, {Yang}, {Chaplin}, {Christensen-Dalsgaard}, {Eggenberger}, {Gizon}, {Mathis}, {Molenda-{\.Z}akowicz}, \& {Pinsonneault}}]{deheuvels+14}
{Deheuvels}, S., {Do{\u{g}}an}, G., {Goupil}, M.~J., {et~al.} 2014, \aap, 564, A27

\bibitem[{{Deheuvels} {et~al.}(2012){Deheuvels}, {Garc{\'\i}a}, {Chaplin}, {Basu}, {Antia}, {Appourchaux}, {Benomar}, {Davies}, {Elsworth}, {Gizon}, {Goupil}, {Reese}, {Regulo}, {Schou}, {Stahn}, {Casagrande}, {Christensen-Dalsgaard}, {Fischer}, {Hekker}, {Kjeldsen}, {Mathur}, {Mosser}, {Pinsonneault}, {Valenti}, {Christiansen}, {Kinemuchi}, \& {Mullally}}]{deheuvels+12}
{Deheuvels}, S., {Garc{\'\i}a}, R.~A., {Chaplin}, W.~J., {et~al.} 2012, \apj, 756, 19

\bibitem[{{Deheuvels} {et~al.}(2023){Deheuvels}, {Li}, {Ballot}, \& {Ligni{\`e}res}}]{deheuvels+23}
{Deheuvels}, S., {Li}, G., {Ballot}, J., \& {Ligni{\`e}res}, F. 2023, \aap, 670, L16

\bibitem[{{Dhouib} {et~al.}(2022){Dhouib}, {Mathis}, {Bugnet}, {Van Reeth}, \& {Aerts}}]{dhouib+22}
{Dhouib}, H., {Mathis}, S., {Bugnet}, L., {Van Reeth}, T., \& {Aerts}, C. 2022, \aap, 661, A133

\bibitem[{{Duez} \& {Mathis}(2010)}]{duez+mathis10}
{Duez}, V. \& {Mathis}, S. 2010, \aap, 517, A58

\bibitem[{{Dupret} {et~al.}(2009){Dupret}, {Belkacem}, {Samadi}, {Montalban}, {Moreira}, {Miglio}, {Godart}, {Ventura}, {Ludwig}, {Grigahc{\`e}ne}, {Goupil}, {Noels}, \& {Caffau}}]{dupret+09}
{Dupret}, M.~A., {Belkacem}, K., {Samadi}, R., {et~al.} 2009, \aap, 506, 57

\bibitem[{{Emeriau-Viard} \& {Brun}(2017)}]{emeriau-viard+17}
{Emeriau-Viard}, C. \& {Brun}, A.~S. 2017, \apj, 846, 8

\bibitem[{{Fuller} {et~al.}(2015){Fuller}, {Cantiello}, {Stello}, {Garcia}, \& {Bildsten}}]{fuller+15}
{Fuller}, J., {Cantiello}, M., {Stello}, D., {Garcia}, R.~A., \& {Bildsten}, L. 2015, Science, 350, 423

\bibitem[{{Garc{\'\i}a} {et~al.}(2014){Garc{\'\i}a}, {P{\'e}rez Hern{\'a}ndez}, {Benomar}, {Silva Aguirre}, {Ballot}, {Davies}, {Do{\u{g}}an}, {Stello}, {Christensen-Dalsgaard}, {Houdek}, {Ligni{\`e}res}, {Mathur}, {Takata}, {Ceillier}, {Chaplin}, {Mathis}, {Mosser}, {Ouazzani}, {Pinsonneault}, {Reese}, {R{\'e}gulo}, {Salabert}, {Thompson}, {van Saders}, {Neiner}, \& {De Ridder}}]{garcia+14}
{Garc{\'\i}a}, R.~A., {P{\'e}rez Hern{\'a}ndez}, F., {Benomar}, O., {et~al.} 2014, \aap, 563, A84

\bibitem[{{Gehan} {et~al.}(2018){Gehan}, {Mosser}, {Michel}, {Samadi}, \& {Kallinger}}]{gehan+18}
{Gehan}, C., {Mosser}, B., {Michel}, E., {Samadi}, R., \& {Kallinger}, T. 2018, \aap, 616, A24

\bibitem[{{Gough} \& {Thompson}(1990)}]{gough&thompson90}
{Gough}, D.~O. \& {Thompson}, M.~J. 1990, \mnras, 242, 25

\bibitem[{{Grosjean} {et~al.}(2014){Grosjean}, {Dupret}, {Belkacem}, {Montalban}, {Samadi}, \& {Mosser}}]{grosjean+14}
{Grosjean}, M., {Dupret}, M.~A., {Belkacem}, K., {et~al.} 2014, \aap, 572, A11

\bibitem[{{Hatt} {et~al.}(2024){Hatt}, {Ong}, {Nielsen}, {Chaplin}, {Davies}, {Deheuvels}, {Ballot}, {Li}, \& {Bugnet}}]{hatt+24}
{Hatt}, E.~J., {Ong}, J.~M.~J., {Nielsen}, M.~B., {et~al.} 2024, \mnras [\eprint[arXiv]{2409.01157}]

\bibitem[{{Hekker} \& {Christensen-Dalsgaard}(2017)}]{hekker+17}
{Hekker}, S. \& {Christensen-Dalsgaard}, J. 2017, \aapr, 25, 1

\bibitem[{{Jermyn} {et~al.}(2023){Jermyn}, {Bauer}, {Schwab}, {Farmer}, {Ball}, {Bellinger}, {Dotter}, {Joyce}, {Marchant}, {Mombarg}, {Wolf}, {Sunny Wong}, {Cinquegrana}, {Farrell}, {Smolec}, {Thoul}, {Cantiello}, {Herwig}, {Toloza}, {Bildsten}, {Townsend}, \& {Timmes}}]{mesa6}
{Jermyn}, A.~S., {Bauer}, E.~B., {Schwab}, J., {et~al.} 2023, \apjs, 265, 15

\bibitem[{{Kallinger} {et~al.}(2014){Kallinger}, {De Ridder}, {Hekker}, {Mathur}, {Mosser}, {Gruberbauer}, {Garc{\'\i}a}, {Karoff}, \& {Ballot}}]{kallinger+14}
{Kallinger}, T., {De Ridder}, J., {Hekker}, S., {et~al.} 2014, \aap, 570, A41

\bibitem[{{Kaufman} {et~al.}(2022){Kaufman}, {Lecoanet}, {Anders}, {Brown}, {Vasil}, {Oishi}, \& {Burns}}]{kaufman+22}
{Kaufman}, E., {Lecoanet}, D., {Anders}, E.~H., {et~al.} 2022, \mnras, 517, 3332

\bibitem[{{Lecoanet} {et~al.}(2022){Lecoanet}, {Bowman}, \& {Van Reeth}}]{lecoanet+22}
{Lecoanet}, D., {Bowman}, D.~M., \& {Van Reeth}, T. 2022, \mnras, 512, L16

\bibitem[{{Lecoanet} {et~al.}(2017){Lecoanet}, {Vasil}, {Fuller}, {Cantiello}, \& {Burns}}]{lecoanet+17}
{Lecoanet}, D., {Vasil}, G.~M., {Fuller}, J., {Cantiello}, M., \& {Burns}, K.~J. 2017, \mnras, 466, 2181

\bibitem[{{Li} {et~al.}(2024){Li}, {Aerts}, {Bedding}, {Fritzewski}, {Murphy}, {Van Reeth}, {Montet}, {Jian}, {Mombarg}, {Gossage}, \& {Sreenivas}}]{li+24}
{Li}, G., {Aerts}, C., {Bedding}, T.~R., {et~al.} 2024, \aap, 686, A142

\bibitem[{{Li} {et~al.}(2022){Li}, {Deheuvels}, {Ballot}, \& {Ligni{\`e}res}}]{li+22}
{Li}, G., {Deheuvels}, S., {Ballot}, J., \& {Ligni{\`e}res}, F. 2022, \nat, 610, 43

\bibitem[{{Li} {et~al.}(2023){Li}, {Deheuvels}, {Li}, {Ballot}, \& {Ligni{\`e}res}}]{li+23}
{Li}, G., {Deheuvels}, S., {Li}, T., {Ballot}, J., \& {Ligni{\`e}res}, F. 2023, \aap, 680, A26

\bibitem[{{Loi}(2020{\natexlab{a}})}]{loi20a}
{Loi}, S.~T. 2020{\natexlab{a}}, \mnras, 493, 5726

\bibitem[{{Loi}(2020{\natexlab{b}})}]{loi20b}
{Loi}, S.~T. 2020{\natexlab{b}}, \mnras, 496, 3829

\bibitem[{{Loi}(2021)}]{loi+21}
{Loi}, S.~T. 2021, \mnras, 504, 3711

\bibitem[{{Loi} \& {Papaloizou}(2017)}]{loi+papaloizou17}
{Loi}, S.~T. \& {Papaloizou}, J. C.~B. 2017, \mnras, 467, 3212

\bibitem[{{Loi} \& {Papaloizou}(2018)}]{loi+papaloizou18}
{Loi}, S.~T. \& {Papaloizou}, J. C.~B. 2018, \mnras, 477, 5338

\bibitem[{Markey \& Tayler(1973)}]{markey+taylor73}
Markey, P. \& Tayler, R.~J. 1973, Monthly Notices of the Royal Astronomical Society, 163, 77

\bibitem[{{Mathis} {et~al.}(2021){Mathis}, {Bugnet}, {Prat}, {Augustson}, {Mathur}, \& {Garcia}}]{mathis+21}
{Mathis}, S., {Bugnet}, L., {Prat}, V., {et~al.} 2021, \aap, 647, A122

\bibitem[{{Mathis} \& {de Brye}(2011)}]{mathis+debrye11}
{Mathis}, S. \& {de Brye}, N. 2011, \aap, 526, A65

\bibitem[{{Mosser} {et~al.}(2011){Mosser}, {Barban}, {Montalb{\'a}n}, {Beck}, {Miglio}, {Belkacem}, {Goupil}, {Hekker}, {De Ridder}, {Dupret}, {Elsworth}, {Noels}, {Baudin}, {Michel}, {Samadi}, {Auvergne}, {Baglin}, \& {Catala}}]{mosser+11}
{Mosser}, B., {Barban}, C., {Montalb{\'a}n}, J., {et~al.} 2011, \aap, 532, A86

\bibitem[{{Mosser} {et~al.}(2017){Mosser}, {Belkacem}, {Pin{\c{c}}on}, {Takata}, {Vrard}, {Barban}, {Goupil}, {Kallinger}, \& {Samadi}}]{mosser+17}
{Mosser}, B., {Belkacem}, K., {Pin{\c{c}}on}, C., {et~al.} 2017, \aap, 598, A62

\bibitem[{{Mosser} {et~al.}(2012{\natexlab{a}}){Mosser}, {Elsworth}, {Hekker}, {Huber}, {Kallinger}, {Mathur}, {Belkacem}, {Goupil}, {Samadi}, {Barban}, {Bedding}, {Chaplin}, {Garc{\'\i}a}, {Stello}, {De Ridder}, {Middour}, {Morris}, \& {Quintana}}]{mosser+12}
{Mosser}, B., {Elsworth}, Y., {Hekker}, S., {et~al.} 2012{\natexlab{a}}, \aap, 537, A30

\bibitem[{{Mosser} {et~al.}(2012{\natexlab{b}}){Mosser}, {Goupil}, {Belkacem}, {Marques}, {Beck}, {Bloemen}, {De Ridder}, {Barban}, {Deheuvels}, {Elsworth}, {Hekker}, {Kallinger}, {Ouazzani}, {Pinsonneault}, {Samadi}, {Stello}, {Garc{\'\i}a}, {Klaus}, {Li}, {Mathur}, \& {Morris}}]{mosser+12b}
{Mosser}, B., {Goupil}, M.~J., {Belkacem}, K., {et~al.} 2012{\natexlab{b}}, \aap, 548, A10

\bibitem[{{Paxton} {et~al.}(2011){Paxton}, {Bildsten}, {Dotter}, {Herwig}, {Lesaffre}, \& {Timmes}}]{mesa1}
{Paxton}, B., {Bildsten}, L., {Dotter}, A., {et~al.} 2011, \apjs, 192, 3

\bibitem[{{Paxton} {et~al.}(2013){Paxton}, {Cantiello}, {Arras}, {Bildsten}, {Brown}, {Dotter}, {Mankovich}, {Montgomery}, {Stello}, {Timmes}, \& {Townsend}}]{mesa2}
{Paxton}, B., {Cantiello}, M., {Arras}, P., {et~al.} 2013, \apjs, 208, 4

\bibitem[{{Paxton} {et~al.}(2015){Paxton}, {Marchant}, {Schwab}, {Bauer}, {Bildsten}, {Cantiello}, {Dessart}, {Farmer}, {Hu}, {Langer}, {Townsend}, {Townsley}, \& {Timmes}}]{mesa3}
{Paxton}, B., {Marchant}, P., {Schwab}, J., {et~al.} 2015, \apjs, 220, 15

\bibitem[{{Paxton} {et~al.}(2018){Paxton}, {Schwab}, {Bauer}, {Bildsten}, {Blinnikov}, {Duffell}, {Farmer}, {Goldberg}, {Marchant}, {Sorokina}, {Thoul}, {Townsend}, \& {Timmes}}]{mesa4}
{Paxton}, B., {Schwab}, J., {Bauer}, E.~B., {et~al.} 2018, \apjs, 234, 34

\bibitem[{{Paxton} {et~al.}(2019){Paxton}, {Smolec}, {Schwab}, {Gautschy}, {Bildsten}, {Cantiello}, {Dotter}, {Farmer}, {Goldberg}, {Jermyn}, {Kanbur}, {Marchant}, {Thoul}, {Townsend}, {Wolf}, {Zhang}, \& {Timmes}}]{mesa5}
{Paxton}, B., {Smolec}, R., {Schwab}, J., {et~al.} 2019, \apjs, 243, 10

\bibitem[{{Prat} {et~al.}(2019){Prat}, {Mathis}, {Buysschaert}, {Van Beeck}, {Bowman}, {Aerts}, \& {Neiner}}]{prat+19}
{Prat}, V., {Mathis}, S., {Buysschaert}, B., {et~al.} 2019, \aap, 627, A64

\bibitem[{{Prat} {et~al.}(2020){Prat}, {Mathis}, {Neiner}, {Van Beeck}, {Bowman}, \& {Aerts}}]{prat+20}
{Prat}, V., {Mathis}, S., {Neiner}, C., {et~al.} 2020, \aap, 636, A100

\bibitem[{{Prendergast}(1956)}]{prendergast56}
{Prendergast}, K.~H. 1956, \apj, 123, 498

\bibitem[{{Rui} \& {Fuller}(2023)}]{rui+fuller23}
{Rui}, N.~Z. \& {Fuller}, J. 2023, \mnras, 523, 582

\bibitem[{{Rui} {et~al.}(2024){Rui}, {Ong}, \& {Mathis}}]{rui+23}
{Rui}, N.~Z., {Ong}, J.~M.~J., \& {Mathis}, S. 2024, \mnras, 527, 6346

\bibitem[{{Shibahashi}(1979)}]{shibahashi79}
{Shibahashi}, H. 1979, \pasj, 31, 87

\bibitem[{{Stello} {et~al.}(2016{\natexlab{a}}){Stello}, {Cantiello}, {Fuller}, {Garcia}, \& {Huber}}]{stello+16a}
{Stello}, D., {Cantiello}, M., {Fuller}, J., {Garcia}, R.~A., \& {Huber}, D. 2016{\natexlab{a}}, \pasa, 33, e011

\bibitem[{{Stello} {et~al.}(2016{\natexlab{b}}){Stello}, {Cantiello}, {Fuller}, {Huber}, {Garc{\'\i}a}, {Bedding}, {Bildsten}, \& {Silva Aguirre}}]{stello+16b}
{Stello}, D., {Cantiello}, M., {Fuller}, J., {et~al.} 2016{\natexlab{b}}, \nat, 529, 364

\bibitem[{{Takata}(2016{\natexlab{a}})}]{takata16a}
{Takata}, M. 2016{\natexlab{a}}, \pasj, 68, 109

\bibitem[{{Takata}(2016{\natexlab{b}})}]{takata16b}
{Takata}, M. 2016{\natexlab{b}}, \pasj, 68, 91

\bibitem[{Tayler(1973)}]{taylor73}
Tayler, R.~J. 1973, Monthly Notices of the Royal Astronomical Society, 161, 365

\bibitem[{Tayler(1980)}]{taylor80}
Tayler, R.~J. 1980, Monthly Notices of the Royal Astronomical Society, 191, 151

\bibitem[{{Unno} {et~al.}(1989){Unno}, {Osaki}, {Ando}, {Saio}, \& {Shibahashi}}]{unno+89}
{Unno}, W., {Osaki}, Y., {Ando}, H., {Saio}, H., \& {Shibahashi}, H. 1989, {Nonradial oscillations of stars}

\bibitem[{{Villebrun} {et~al.}(2019){Villebrun}, {Alecian}, {Hussain}, {Bouvier}, {Folsom}, {Lebreton}, {Amard}, {Charbonnel}, {Gallet}, {Haemmerl{\'e}}, {B{\"o}hm}, {Johns-Krull}, {Kochukhov}, {Marsden}, {Morin}, \& {Petit}}]{villebrun+19}
{Villebrun}, F., {Alecian}, E., {Hussain}, G., {et~al.} 2019, \aap, 622, A72

\bibitem[{{Yu} {et~al.}(2018){Yu}, {Huber}, {Bedding}, {Stello}, {Hon}, {Murphy}, \& {Khanna}}]{yu+18}
{Yu}, J., {Huber}, D., {Bedding}, T.~R., {et~al.} 2018, \apjs, 236, 42

\end{thebibliography}

\begin{appendix}

\section{Additional information on the magneto-gravity ray tracing} 

\subsection{Buoyancy frequency profiles} \label{app: buoyancy frequency profiles}

\begin{figure}[h!]
    \sidecaption
    \resizebox{\hsize}{!}{\includegraphics{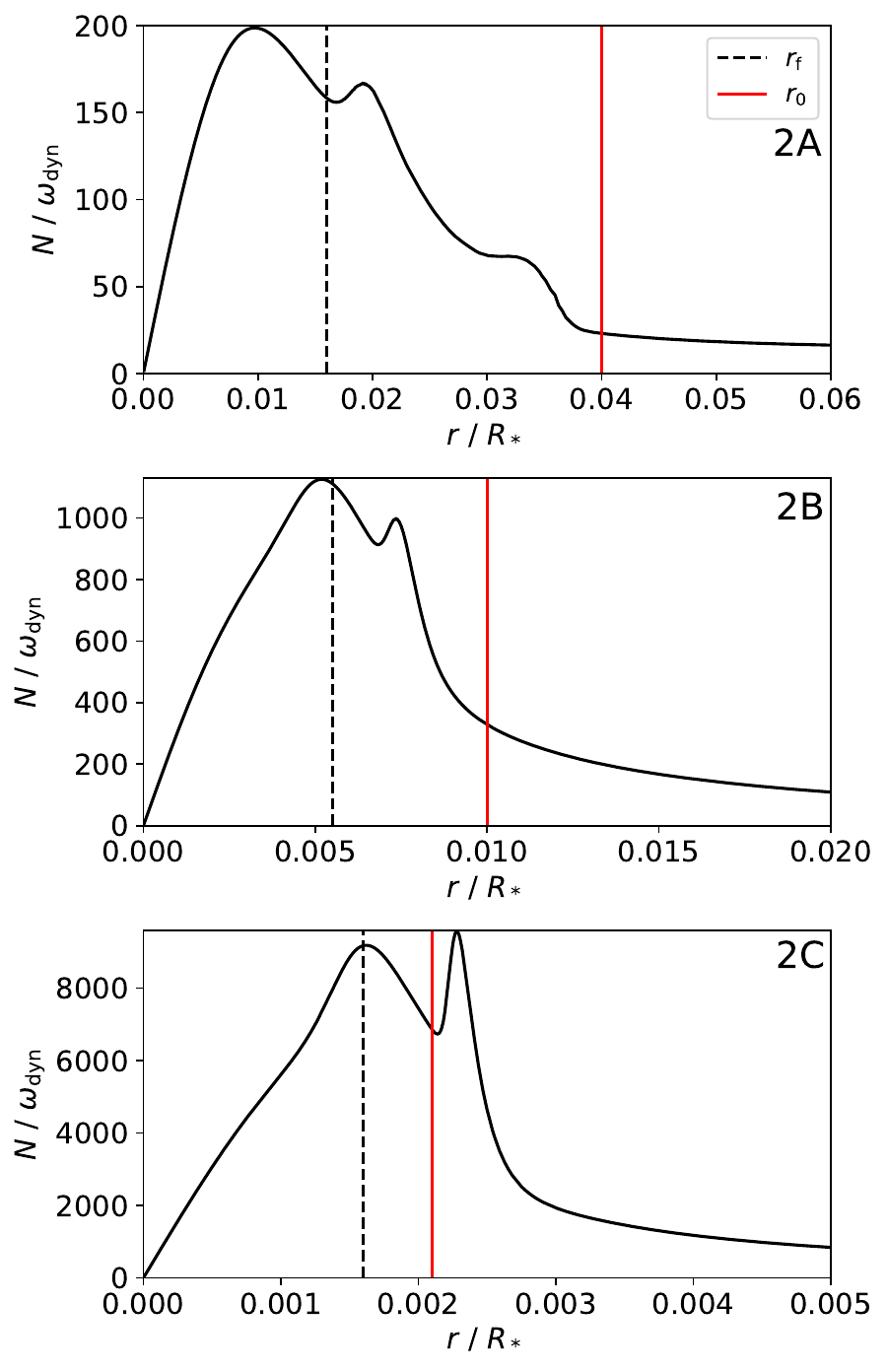}}
    \caption{Profiles of the buoyancy frequency as a function of radius for the models 2A, 2B, and 2C. The launch radius of the rays $r_0$ is shown in red for each model, the maximum extend of the internal magnetic field $r_{\rm f}$ is shown as the dashed black line.}
    \label{fig: N profiles}
\end{figure}

In Fig. \ref{fig: N profiles}, we show the buoyancy frequency profiles of the models 2A, 2B, and 2C. As the models were selected to have a similar radius to models A, B and C of \citet{loi20a}, the overall qualitative trends remain unchanged compared to these models. However, there are small quantitative differences between our buoyancy frequency profiles and the profiles of \citet{loi20a} due to the changes in the MESA version, the inlist, and the interpolation onto another grid performed by \citet{loi20a}.

\subsection{Hamilton's equations for magneto-gravity waves} \label{app: Equations Hamiltonian}

Here, we show the set of first-order ordinary differential equations describing the propagation of a magneto-gravity wave packet in time in spherical polar coordinates \citep{loi20a}:
\begin{gather}
    \frac{{\rm d}r}{{\rm d}t} = \frac{\omega_{\rm A}}{\omega} V_{{\rm A}r} - \frac{N^2 \kappa^2_\perp \kappa_r}{\omega k}, 
    \label{eq: hamiltion r}\\
    \frac{{\rm d}\theta}{{\rm d}t} = \frac{\omega_{\rm A}}{\omega} \frac{V_{{\rm A}\theta}}{r} + \frac{N^2 \kappa_{\theta} \kappa^2_r}{\omega k r}, \\
    \frac{{\rm d}\phi}{{\rm d}t} = \frac{\omega_{\rm A}}{\omega} \frac{V_{{\rm A}\phi}}{r \sin\theta} + \frac{N^2 \kappa_{\phi} \kappa^2_r}{\omega k r \sin\theta}, \\
    \frac{{\rm d}k_r}{{\rm d}t} = \frac{\omega_{\rm A}}{\omega} \left( \frac{k_\theta V_{{\rm A}\theta} + k_\phi V_{{\rm}\phi}}{r} -  \boldsymbol{\rm k} \cdot \frac{\partial {\boldsymbol{\rm V}}_{\rm A}}{\partial r} \right) \notag\\
        \qquad + \frac{N \kappa^2_\perp}{\omega} \left( \frac{N \kappa^2_r}{r} - \frac{{\rm d}N}{{\rm d}r} \right), \\
    \frac{{\rm d}k_\theta}{{\rm d}t} = - \frac{\omega_{\rm A}}{\omega} \frac{1}{r} \left( k_\theta V_{{\rm A}r} - k_\phi V_{{\rm A}\phi} \cot\theta + \boldsymbol{\rm k} \cdot \frac{\partial {\boldsymbol{\rm V}}_{\rm A}}{\partial \theta} \right) \notag\\
        \qquad + \frac{N^2 \kappa_r}{r \omega} (\kappa_\theta \kappa^2_\perp + \kappa^2_\phi \kappa_r \cot\theta), \\
    \frac{{\rm d}k_\phi}{{\rm d}t} = - \frac{\omega_{\rm A}}{\omega} \frac{k_\phi}{r} (V_{{\rm A}r} + V_{{\rm A}\theta} \cot\theta) + \frac{N^2 \kappa_\phi \kappa_r}{r \omega} (\kappa^2_\perp - \kappa_\theta \kappa_r \cot\theta). \label{eq: hamilton k_phi}
\end{gather}
In these equations, $\boldsymbol{\kappa} = (\kappa_r,\kappa_\theta,\kappa_\phi) = \boldsymbol{\rm k}/k$ and $\kappa_\perp = \sqrt{\kappa^2_\theta + \kappa^2_\phi}$.

\subsection{Correction of the wavevector} \label{app: adjustment wavevector}

As described in Sect. \ref{sect: ray initialization}, we adjusted the radial component of the wavevector $k_r$ at each timestep to conserve $\omega$ during the ray tracing simulation. This was achieved by finding the roots of the following equation which can be derived from the dispersion relation \citep[][]{loi20a}:
\begin{gather}
    0 = b_1 k^4_r + b_2 k^3_r + b_3 k^2_r + b_4 k_r + b_5, 
    \label{eq: fix k_r 1}
\end{gather}
with
\begin{gather}
    b_1 = V^2_{{\rm A}r}, \\
    b_2 = 2 V_{{\rm A}r} (k_\theta V_{{\rm A}\theta} + k_\phi V_{{\rm A}\phi}), \\
    b_3 = V^2_{{\rm A}r} k^2_\perp + (k_\theta V_{{\rm A}\theta} + k_\phi V_{{\rm A}\phi})^2 - \omega^2, \\
    b_4 = 2 V_{{\rm A}r} k^2_\perp (k_\theta V_{{\rm A}\theta} + k_\phi V_{{\rm A}\phi}), \\
    b_5 = k^2_\perp [(k_\theta V_{{\rm A}\theta} + k_\phi V_{{\rm A}\phi})^2 + N^2 - \omega^2].
\end{gather}

\citet{loi20a} used a Newton-Raphson iteration scheme to find the roots of Eq. \ref{eq: fix k_r 1}. They performed a maximum of five iterations after each timestep during the integration of each ray. If the scheme has not converged after five iterations, they set the value of $k_r$ to the current estimate of the root and continued the integration of the ray. In this case, $\omega$ deviates from its initial value. Thus, the value of $\omega$ after each timestep can be used as a measure of accuracy. 

Instead of using a Newton-Raphson scheme, we directly calculated the roots and selected the one with the real part closest to the current estimate of $k_r$ as the new value of $k_r$ (see Sect. \ref{sect: ray initialization} for details). Since we have discarded the imaginary part, $\omega$ can also be used as a measure of accuracy for our method. 
We show a comparison of a set of 1200 rays using the Newton-Raphson method to correct $k_r$ with a set of rays where we used our method to correct $k_r$ in Appendix \ref{sect: setup comparison}.
Overall, we find that calculating the roots directly improves the stability of the integration and reduces the uncertainties.

\subsection{Ray classification} \label{app: ray classification}

Since it is not feasible to check each ray manually, we used a ray classification scheme to decide whether a ray is trapped, reflected, or unclassified. 
It consists of a series of conditions that are checked in sequence. If the ray fulfilled a condition, it was classified accordingly and the other steps were ignored. If the ray did not meet the condition, the next one was checked and so on.
The ray classification scheme used in this work is an updated version of the one used by \citet{loi20a}. It works as follows:
\begin{enumerate}
    \item The value of $\omega$ is different than the original value by more than 0.0001 \% at any point during the integration. \\
    $\rightarrow$ Unclassified
    \item The ray crosses $r_{\rm f}$ going outward.
    \begin{enumerate}
        \item The final value of $k_r$ is equal or larger than 1000 times the initial value of $k_r$\footnote{If this is the case, we find that $k_r$ is not bounded, which is a key property of reflected rays. Therefore, we identify these rays as unclassified.}. \\
        $\rightarrow$ Unclassified
        \item The final value of $k_r$ is smaller than 1000 times the initial value of $k_r$. \\
        $\rightarrow$ Reflected
    \end{enumerate}
    \item The final value of $k_r$ exceeds the initial value $k_r$ by a factor of 10.
    \begin{enumerate}
        \item The initial value of $k_r$ is not always smaller than the value of $k_r$ during the second half of the integration\footnote{If this is the case, $k_r$ does not diverge, which is a key property of trapped rays. Therefore, we identify these rays as unclassified.}. \\
        $\rightarrow$ Unclassified
        \item The initial value of $k_r$ is always smaller than the values of $k_r$ during the second half of the integration. \\
        $\rightarrow$ Trapped
    \end{enumerate}
    \item None of the above conditions is fulfilled.\\
    $\rightarrow$ Unclassified
\end{enumerate}

We have introduced additional quality checks to the ray classification scheme of \citet{loi20a} before identifying a ray as reflected or trapped (see Fig. \ref{fig: unispheres newton}). 
This was done because there were rays with clear numerical artifacts that were classified as either reflected or trapped by the classification scheme of \citet{loi20a}. The majority of these rays were identified as unclassified by our scheme.
Due to the unpredictability of the numerical artifacts, we cannot exclude that there still is a small number of rays that exhibit unphysical behavior and were classified as either trapped or reflected. This may lead to a slight underestimation of the uncertainties of the trapped fraction, but should not have a significant impact on the results presented in this work.
Any ray that was identified as unclassified according to the classification scheme of \citet{loi20a} was also unclassified according to our scheme.

\subsection{Comparison of the ray tracing setups} \label{sect: setup comparison}

\begin{figure}[]
    \sidecaption
    \resizebox{\hsize}{!}{\includegraphics{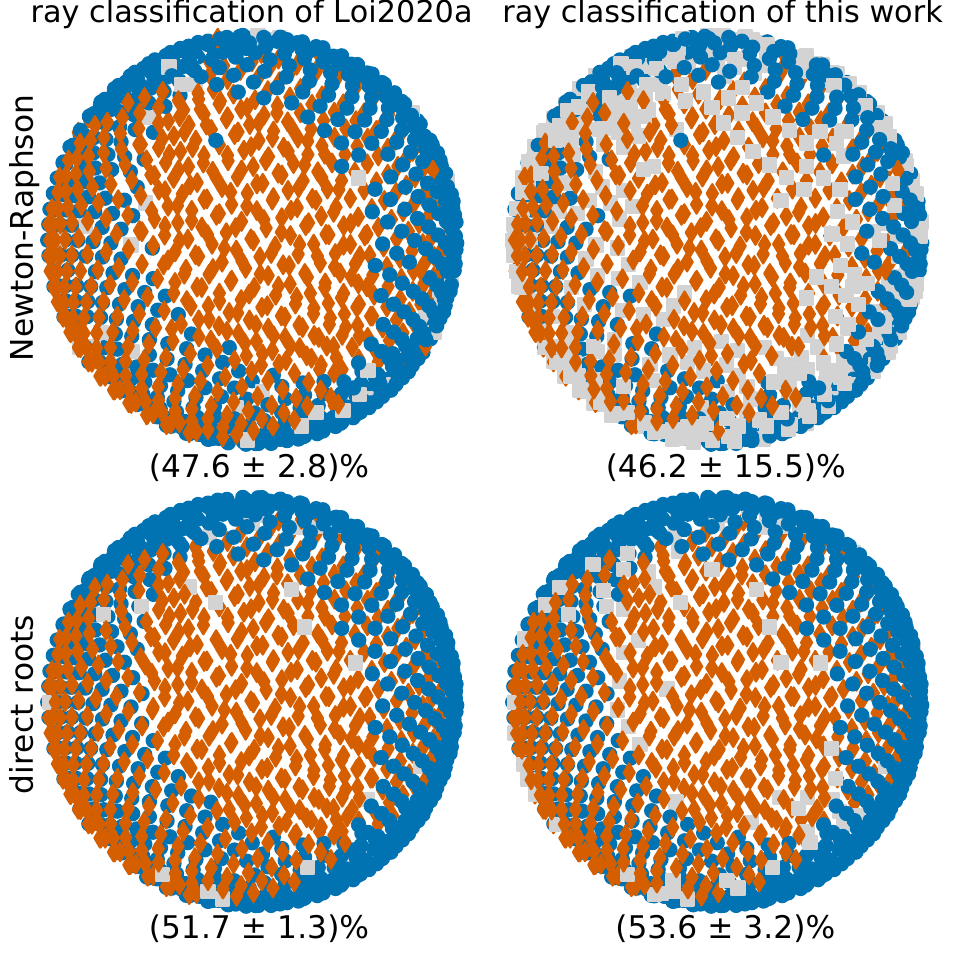}}
    \caption{Sets of 1200 rays plotted as as unispheres parameterized by $(\theta, \phi)  = (\theta_0, \alpha)$ for the model 2A with $B_{\rm cen} = B_{\rm crit,cen}$, $\omega = 10\ \omega_{\rm dyn}$, and $l = 1$. Trapped rays are shown as blue dots, reflected rays are shown as red diamonds, and unclassified rays are shown as gray squares. In the \textit{top row}, we show rays for which we have used a Newton-Raphson iteration scheme for the correction of $k_r$, which was used by \citet{loi20a}. In the \textit{bottom row}, we show rays where we have used our method described in Sect. \ref{sect: ray initialization} to correct $k_r$. In the \textit{left column}, we employed the ray classification scheme of \citet{loi20a} to decide whether a ray is trapped, reflected, or unclassified. In the \textit{right column}, we used our ray classification scheme described in Appendix \ref{app: ray classification}. The trapped fraction $f_{\rm T}$ of each set is indicated under its corresponding unisphere.}
    \label{fig: unispheres newton}
\end{figure}

In Fig. \ref{fig: unispheres newton} we compare a set of 1200 rays using a Newton-Raphson iteration scheme to correct $k_r$ with a set of rays where we used our method to correct $k_r$ (i.e., calculating the roots of Eq. \ref{eq: fix k_r 1} directly; see Sect. \ref{sect: ray initialization} and Appendix \ref{app: adjustment wavevector}). For both sets of rays, we show the ray classification according to the scheme of \citet{loi20a}, as well as according to our ray classification scheme (see Appendix \ref{app: ray classification}). 
Comparing the two classification schemes, the number of unclassified rays is higher when using our classification scheme than when using that of \citet{loi20a}.
Furthermore, we find that a small number of rays that clearly exhibit numerical artifacts were not identified as unclassified when using the classification of \citet{loi20a}. Our classification scheme seems to be more reliable in finding these numerical artifacts. We therefore conclude that the uncertainties of the trapped fraction (Eq. \ref{eq: trapped fraction}) might be underestimated when using the classification scheme of \citet{loi20a}.

The comparison of the two methods for correcting $k_r$ shows that there is more scatter in the behavior of the rays when using the Newton-Raphson iteration than when using our method to determine the roots of Eq. \ref{eq: fix k_r 1}. This can be recognized by the fact that in the sets of rays corresponding to the Newton-Raphson iteration, there is a small number of trapped rays that are completely surrounded by reflected rays and vice versa. There are no such rays when using our method to correct $k_r$.
Nevertheless, the trapped fraction estimated using the Newton-Raphson iteration is not significantly different from the trapped fraction determined using our method when our ray classification scheme is utilized.
In this case however, the uncertainties of the trapped fraction are an order of magnitude smaller when using our method (i.e., calculating the roots of Eq. \ref{eq: fix k_r 1} directly), which is why we applied it in this work.

\subsection{Comparison to \citet{loi20a} and \citet{loi20b}} \label{app: comparison to loi}

\subsubsection{Comparison to \citet{loi20a}}

\begin{figure}[]
    \sidecaption
    \resizebox{\hsize}{!}{\includegraphics{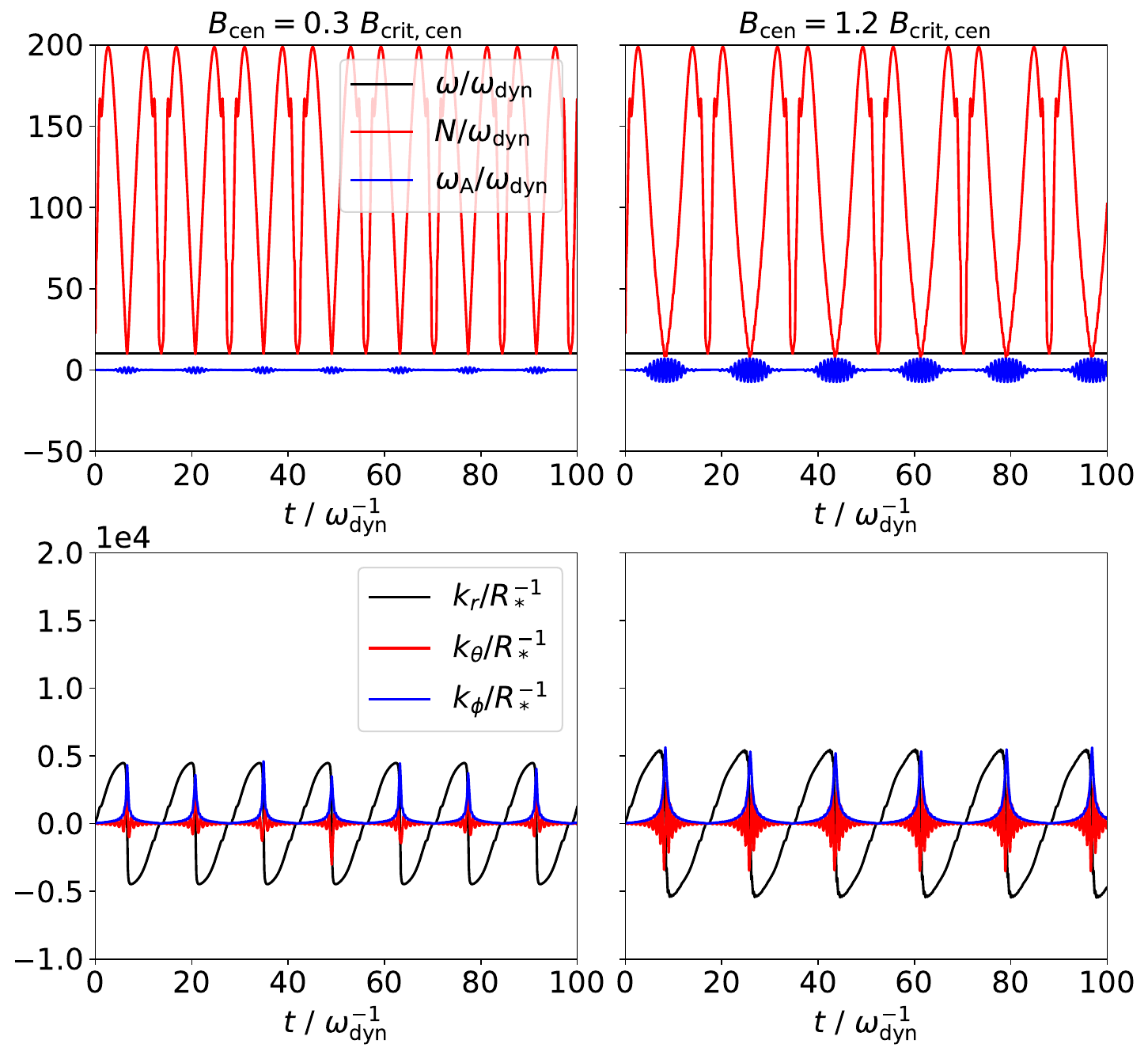}}
    \caption{Reproduction of the lower two panels of Figure 4 of \citet{loi20a} using our implementation of the ray tracing described in Sect. \ref{sect: Ray tracing} for the model 2A. The limits of the axes are also similar to Figure 4 of \citet{loi20a}. In the \textit{left column}, we show the behavior of the ray with $B_{\rm cen} = 0.3\ B_{\rm crit,cen}$, which is the value used by \citet{loi20a}. In the \textit{right column}, we show how the ray behaves with $B_{\rm cen} = 1.2\ B_{\rm crit,cen}$. This value has been selected such that the ray undergoes the same number of reflections as the ray shown in Figure 4 of \citet{loi20a}.
    The remaining initial parameters are identical to \citet{loi20a} (i.e., $\theta_0 = 99^{\circ}$, $\alpha = 40.5^{\circ}$, $\omega = 10\ \omega_{\rm dyn}$, and $l = 1$).}
    \label{fig: Loi20a k components reflected}
\end{figure}

\begin{figure}[]
    \sidecaption
    \resizebox{\hsize}{!}{\includegraphics{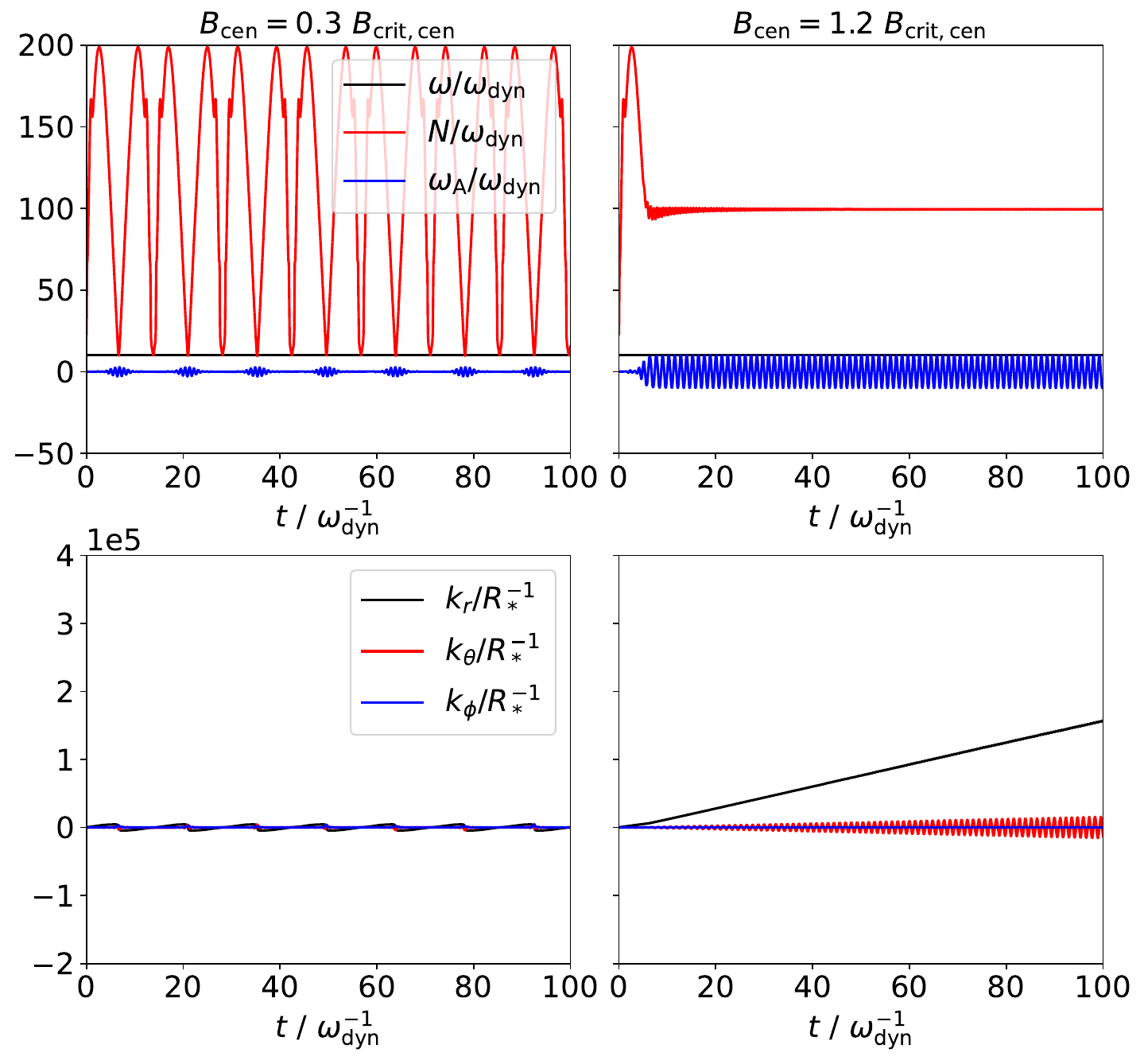}}
    \caption{Reproduction of the lower two panels of Figure 5 of \citet{loi20a} using the same setup as for Fig. \ref{fig: Loi20a k components reflected}. Here, $\theta_0 =  141^{\circ}$ and $\alpha = 58.5^{\circ}$. The remaining initial parameters are the same as for Fig. \ref{fig: Loi20a k components reflected}.}
    \label{fig: Loi20a k components trapped}
\end{figure}

\begin{figure}[]
    \centering
    \resizebox{0.7\hsize}{!}{\includegraphics{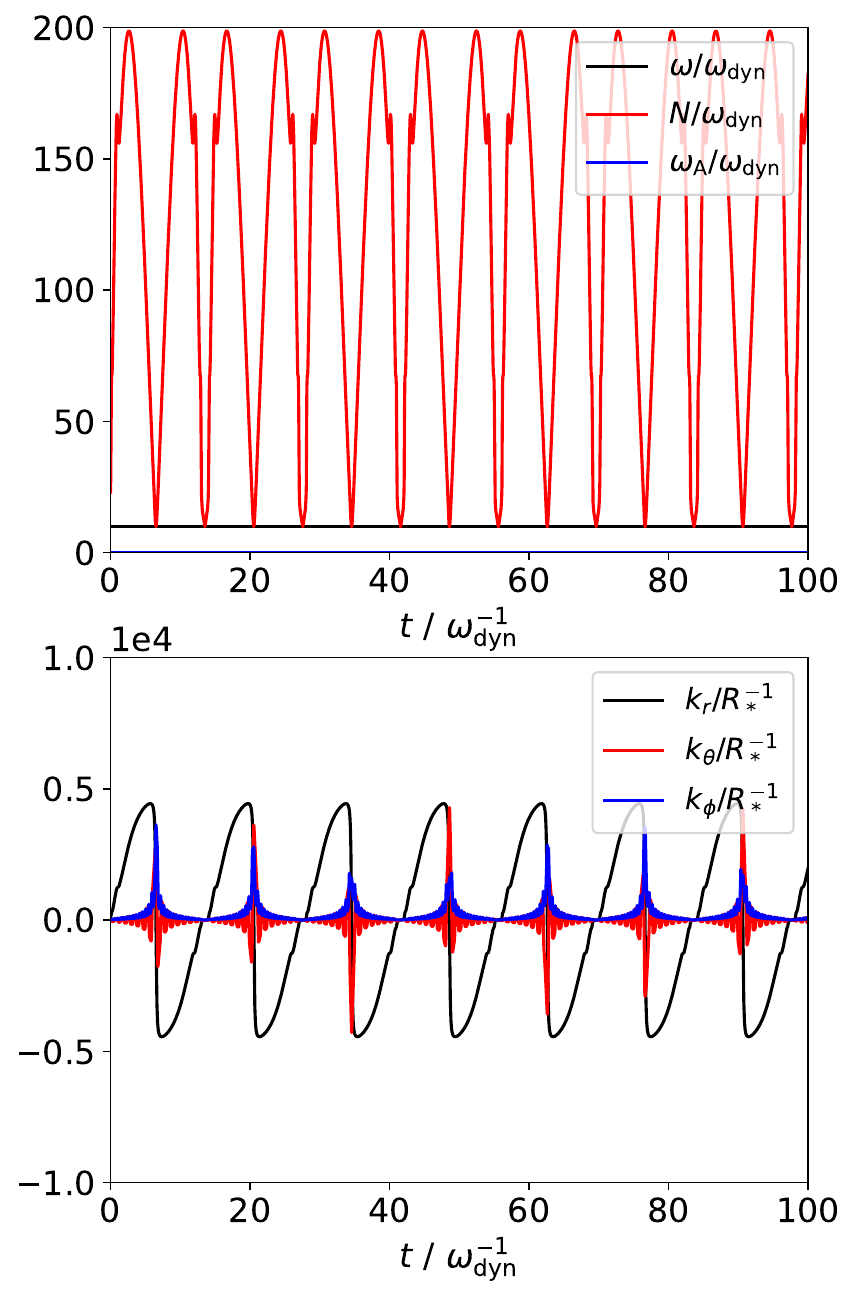}}
    \caption{Reproduction of the lower two panels of Figure 6 of \citet{loi20a} for the ray shown in Fig. \ref{fig: Loi20a k components trapped}, but without an internal magnetic field (i.e., $B_{\rm cen} = 0$).}
    \label{fig: Loi20a k components non-magnetic}
\end{figure}

The numerical integration of the magneto-gravity ray trajectories performed in this work is based on the study of \citet{loi20a}. We used the same set of differential equations (Eqs. \ref{eq: hamiltion r} to \ref{eq: hamilton k_phi}), as well as the same ray initialization and a similar forward Runge-Kutta integration scheme. Therefore, we expect to obtain results consistent with \citet{loi20a} for our models 2A, 2B, and 2C, which are reproductions of their models A, B, and C. While the trapped fraction qualitatively shows the same trends with $B_{\rm cen}$, $\omega$, and $l$ as reported by \citet{loi20a}, we systematically obtain smaller trapped fractions for the same $B_{\rm cen}$.  If for example $B_{\rm cen} = B_{\rm crit,cen}$, we obtain a trapped fraction of roughly 50 \%, while \citet{loi20a} obtains trapped fractions $\gtrsim$ 75 \%.
We find that neither the MESA profiles with increased spacial resolution (see Appendix \ref{app: buoyancy frequency profiles}) nor the changes to the correction of $k_r$ during the integration (see Appendix \ref{app: adjustment wavevector}) are responsible for this discrepancy. The difference in the critical field strengths of our model 2A and model A of \citet{loi20a} (see Table \ref{tab: MESA models}) does not make up for this either.

Since the discrepancy in the trapped fraction is present for all models and combinations of $B_{\rm cen}$, $\omega$, and $l$, it must be caused by a systematic difference of the behavior of the individual rays in our implementation and that of \citet{loi20a}. In Figs. \ref{fig: Loi20a k components reflected} and \ref{fig: Loi20a k components trapped}, we show reproductions of Figures 4 and 5 of \citet{loi20a} using our implementation of the magneto-gravity ray tracing. In addition to the magnetic field strength used by \citet{loi20a} (i.e., $B_{\rm cen} = 0.3\ B_{\rm crit,cen}$), we also show the behavior of the ray for $B_{\rm cen} = 1.2\ B_{\rm crit,cen}$, for which the ray in Fig. \ref{fig: Loi20a k components reflected} undergoes the same number of reflections as the ray shown in Figure 4 of \citet{loi20a}.

In Fig. \ref{fig: Loi20a k components reflected}, we show that we need a magnetic field strength that is about four times greater than that of \citet{loi20a} so that our ray undergoes the same number of reflections as the ray of \citet{loi20a}. Likewise, our ray shown in Fig. \ref{fig: Loi20a k components trapped} is not trapped when using the field strength of \citet{loi20a}, although they find that the ray is trapped. Using a field strength that is four times greater, the ray shown in Fig. \ref{fig: Loi20a k components trapped} is trapped, but its behavior is still significantly different to that of the ray shown in Figure 5 of \citet{loi20a}. This suggests that the discrepancy cannot be explained by an offset in $B_{\rm crit,cen}$. 

Figure 4 of \citet{loi20a} shows a peculiarity that indicates that there are some systematic differences between their and our numerical integration of the rays. In the third panel where \citet{loi20a} shows the evolution of the wavevector components in time, the maximum absolute value of $k_\phi$ is significantly larger than the maximum absolute value of $k_r$. During the propagation of our rays, the maximum absolute value of $k_\phi$ is always comparable or smaller than the maximum absolute value of $k_r$ (see bottom panel of Figs. \ref{fig: Loi20a k components reflected} and \ref{fig: Loi20a k components trapped}). 

In Fig. \ref{fig: Loi20a k components non-magnetic}, we show a reproduction of Figure 6 of \citet{loi20a}. Here, the internal magnetic field has been turned off (i.e., $B_{\rm cen} = 0$). In their Figure 6, the maximum absolute value of $k_\phi$ is larger than the maximum absolute value of $k_r$, similar as in their Figure 4. The same is true for the maximum absolute value of $k_\theta$. This is never the case with our rays (see bottom panel of Fig. \ref{fig: Loi20a k components non-magnetic}). Since the ray tracing equations (Eqs. \ref{eq: hamiltion r} to \ref{eq: hamilton k_phi}) depend on the current value of $k_\theta$ and $k_\phi$, different values of $k_\theta$ and $k_\phi$ can influence the local behavior and thus the trajectory of the ray, as well as its eventual fate (i.e., trapped vs. reflected).

\subsubsection{Comparison to \citet{loi20b}}

\begin{figure}[]
    \sidecaption
    \resizebox{\hsize}{!}{\includegraphics{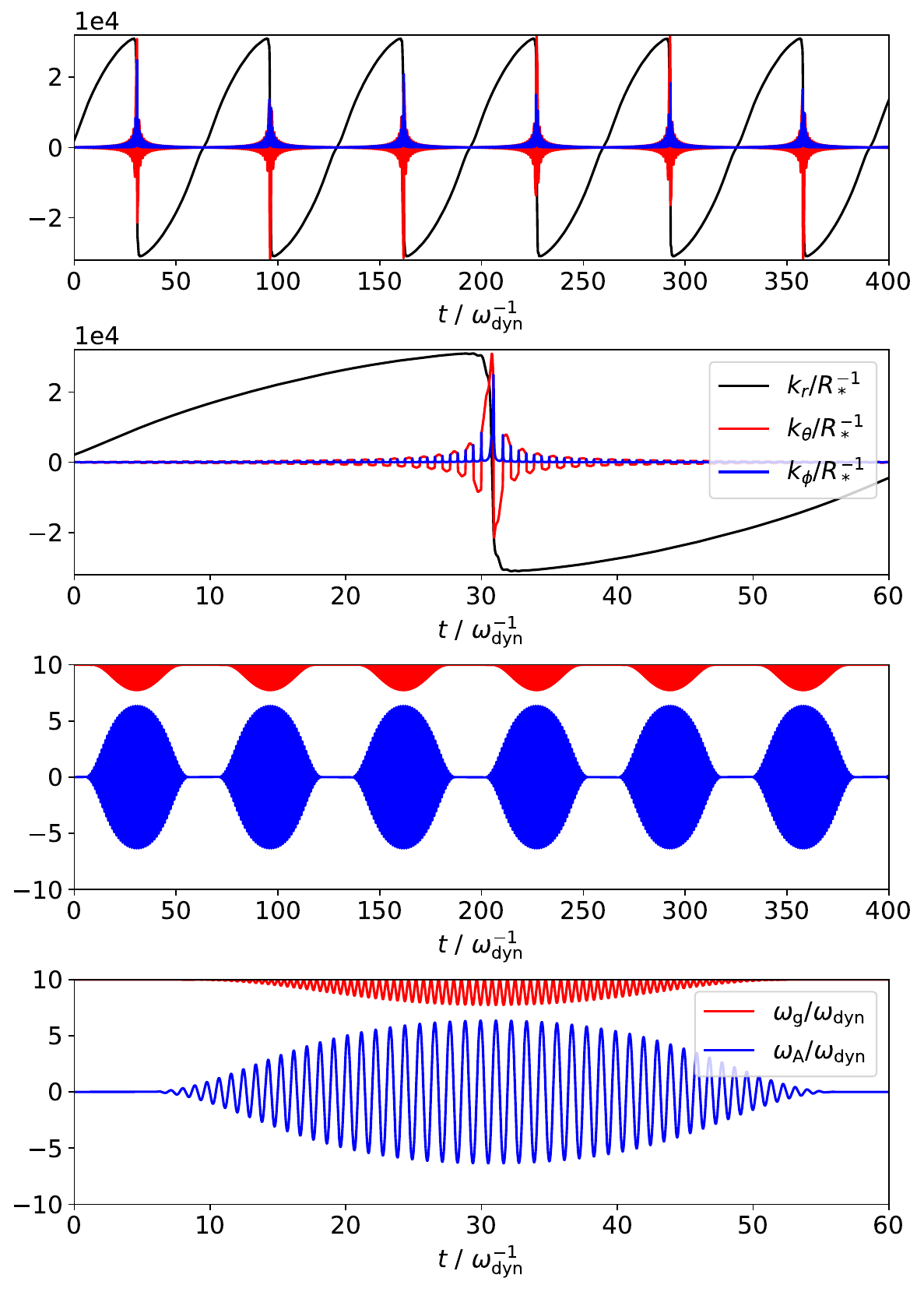}}
    \caption{Reproduction of Figure 4 of \citet{loi20b} using our implementation of the ray tracing described in Sect. \ref{sect: Ray tracing} but with synthetic $N$ and $\rho$ profiles (see Appendix \ref{app: comparison to loi} for details). Since $\theta_0$ and $\alpha$ are not stated by \citet{loi20b}, they have been chosen manually ($\theta_0 = 11.0^{\circ}$ and $\alpha = 81.0^{\circ}$). The remaining initial parameters are identical to \citet{loi20b} (i.e., $r_0 = 0.04\ R_*$, $r_{\rm f} = 0.01\ R_*$, $B_{\rm cen} = B_{\rm crit,cen}$, $\omega = 10\ \omega_{\rm dyn}$, and $l = 1$).}
    \label{fig: Loi20b k components magnetic}
\end{figure}

\begin{figure}[]
    \sidecaption
    \resizebox{\hsize}{!}{\includegraphics{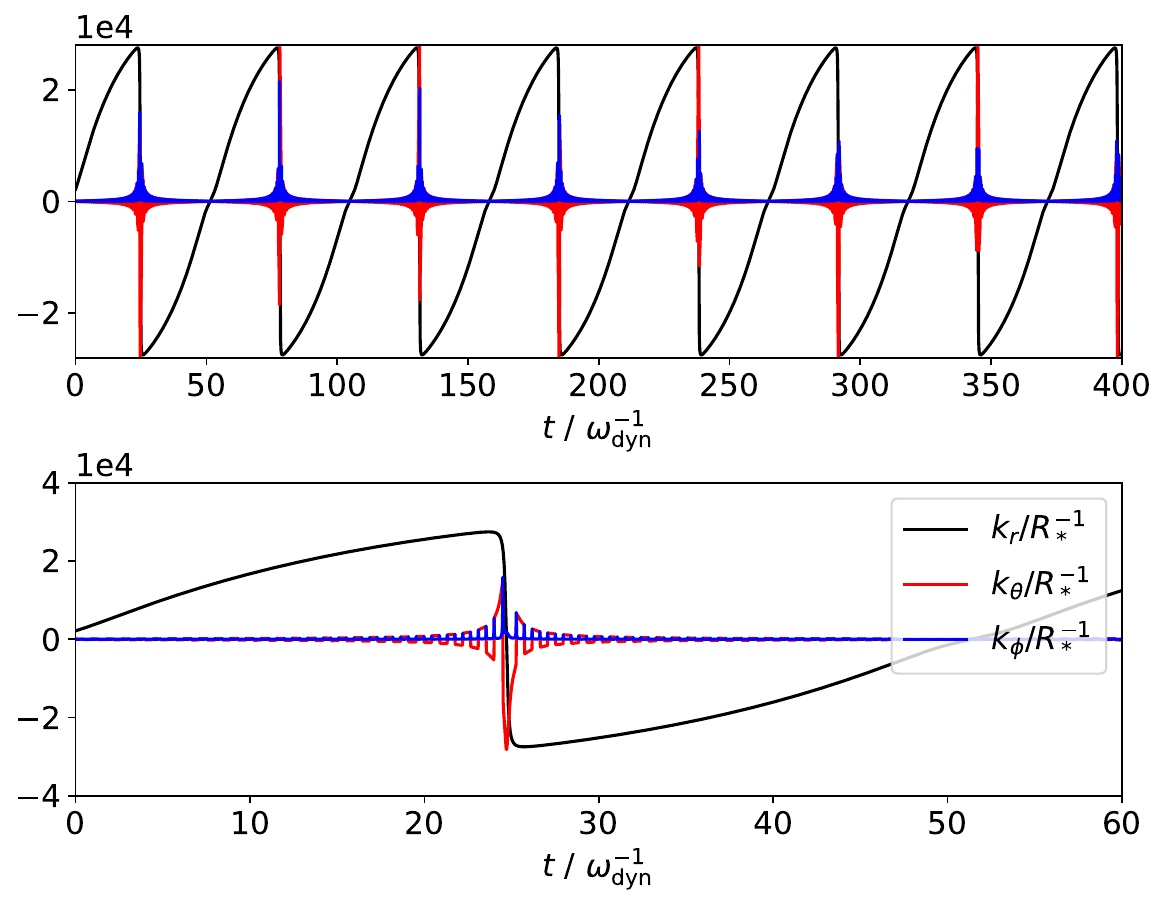}}
    \caption{Reproduction of Figure 5 of \citet{loi20b} for the ray shown in Fig. \ref{fig: Loi20b k components magnetic}, but without an internal magnetic field (i.e., $B_{\rm cen} = 0$).}
    \label{fig: Loi20b k components non-magnetic}
\end{figure}

The magneto-gravity ray tracing method presented by \citet{loi20a} has also been used by \citet{loi20b}. It is a follow-up work of \citet{loi20a} where the author investigates the effect of an internal magnetic field on the period spacing. The only difference to the ray tracing analysis presented by \citet{loi20a} is that \citet{loi20b} performed the integration of the rays on an analytic background instead of a finite grid. Since we want to verify our procedure, which is based on MESA profiles consisting of a finite radial grid, we projected the analytic expressions of \citet{loi20b} for $N$ and $\rho$ on a radial logarithmic grid with 5000 grid cells. We used the resulting grids as mock MESA profiles to compare the behavior of our rays to \citet{loi20b}.

In Figs. \ref{fig: Loi20b k components magnetic} and \ref{fig: Loi20b k components non-magnetic}, we show reproductions of Figures 4 and 5 of \citet{loi20b} using our implementation of the magneto-gravity ray tracing. Since $\theta_0$ and $\alpha$ are not stated by \citet{loi20b}, we chose them manually. Both Figs. \ref{fig: Loi20b k components magnetic} and \ref{fig: Loi20b k components non-magnetic} have a striking resemblance to Figures 4 and 5 of \citet{loi20b} respectively. In particular, the maximum absolute values of $k_\theta$ and $k_\phi$ are always comparable or smaller than the maximum absolute value of $k_r$. This is in contrast to the figures shown by \citet{loi20a}.

In conclusion, the trajectories of our rays are consistent with the most recent implementation of the author of \citet{loi20a}. We hypothesize that the difference in the trapped fraction of \citet{loi20a} and this work can be explained by the systematically different behavior of $k_\theta$ and $k_\phi$. The reason for this discrepancy in the behavior of the wavevector components remains elusive.

\section{Additional figures} \label{app: additional plot ray tracing}

\subsection{Trapped fractions of the remaining models}

\begin{figure}[]
    \sidecaption
    \resizebox{\hsize}{!}{\includegraphics{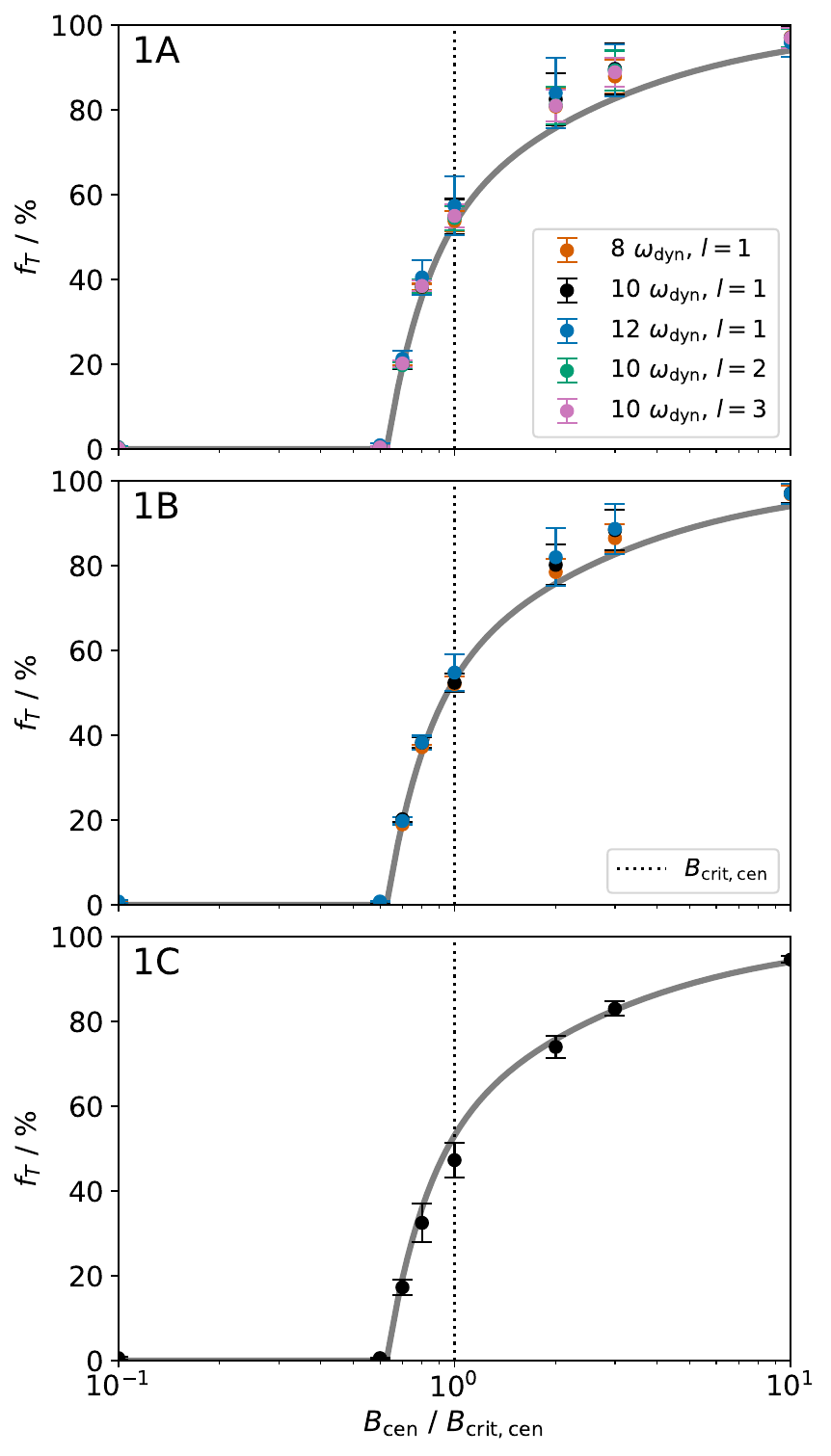}}
    \caption{Same as Fig. \ref{fig: trapped fraction model 2 Bcrit}, now for the models 1A, 1B, and 1C.}
    \label{fig: trapped fraction model 1 Bcrit}
\end{figure}

\begin{figure}[]
    \sidecaption
    \resizebox{\hsize}{!}{\includegraphics{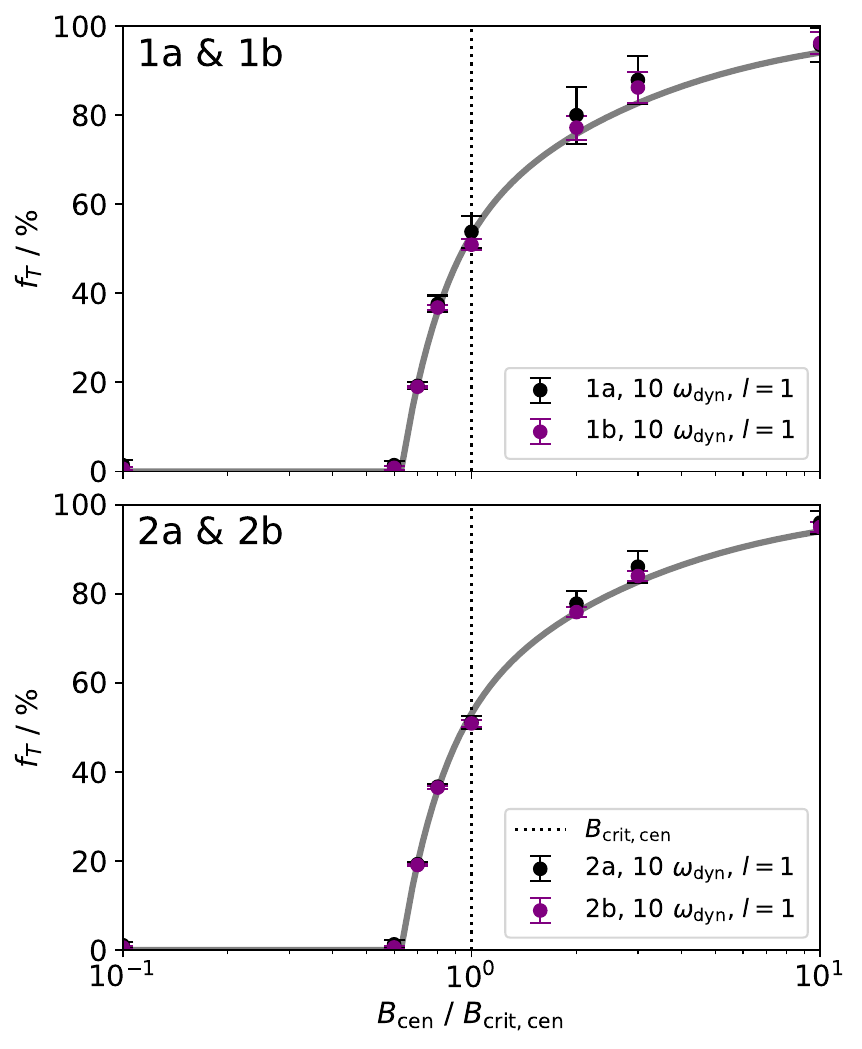}}
    \caption{Same as Fig. \ref{fig: trapped fraction model 2 Bcrit}, now for the models 1a, 1b, 2a, and 2b.}
    \label{fig: trapped fraction model lowZ Bcrit}
\end{figure}

\begin{figure}[]
    \sidecaption
    \resizebox{\hsize}{!}{\includegraphics{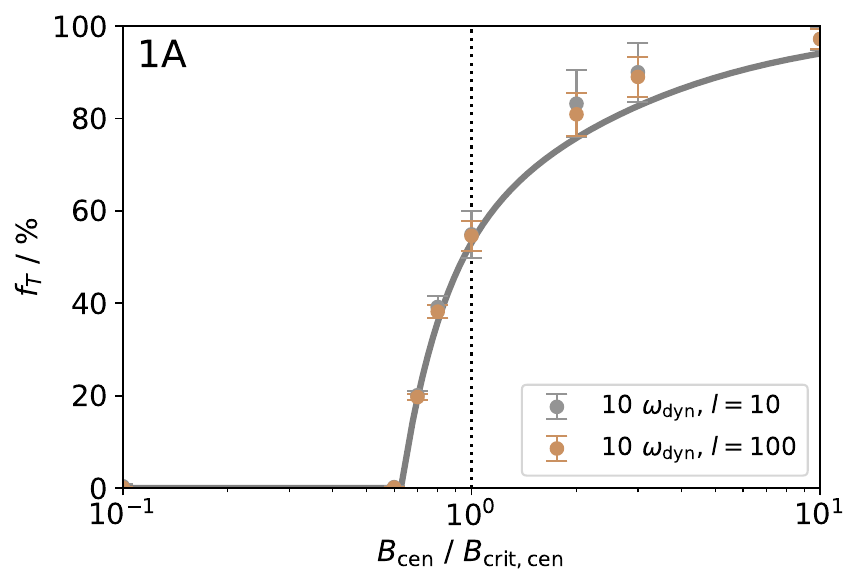}}
    \caption{Same as Fig. \ref{fig: trapped fraction model 2 Bcrit}, now for the model 1A with $\omega = 10\ \omega_{\rm dyn}$ and $l=10$ and 100.}
    \label{fig: trapped fraction l=10 l=100}
\end{figure}

In Figs. \ref{fig: trapped fraction model 1 Bcrit} and \ref{fig: trapped fraction model lowZ Bcrit}, we show the trapped fractions for the models 1A, 1B, and 1C, as well as 1a, 1b, 2a, and 2b. They follow the same trends as the trapped fractions shown in Figs. \ref{fig: trapped fraction model 2} and \ref{fig: trapped fraction model 2 Bcrit}.
In addition, we shown in Fig. \ref{fig: trapped fraction l=10 l=100} the trapped fractions corresponding to a spherical degree of $l = 10$ and $100$ for model 1A. They also follow the same trends.

\subsection{Evolution of the trapped fraction}

\begin{figure*}[]
    \sidecaption
    \resizebox{\hsize}{!}{\includegraphics{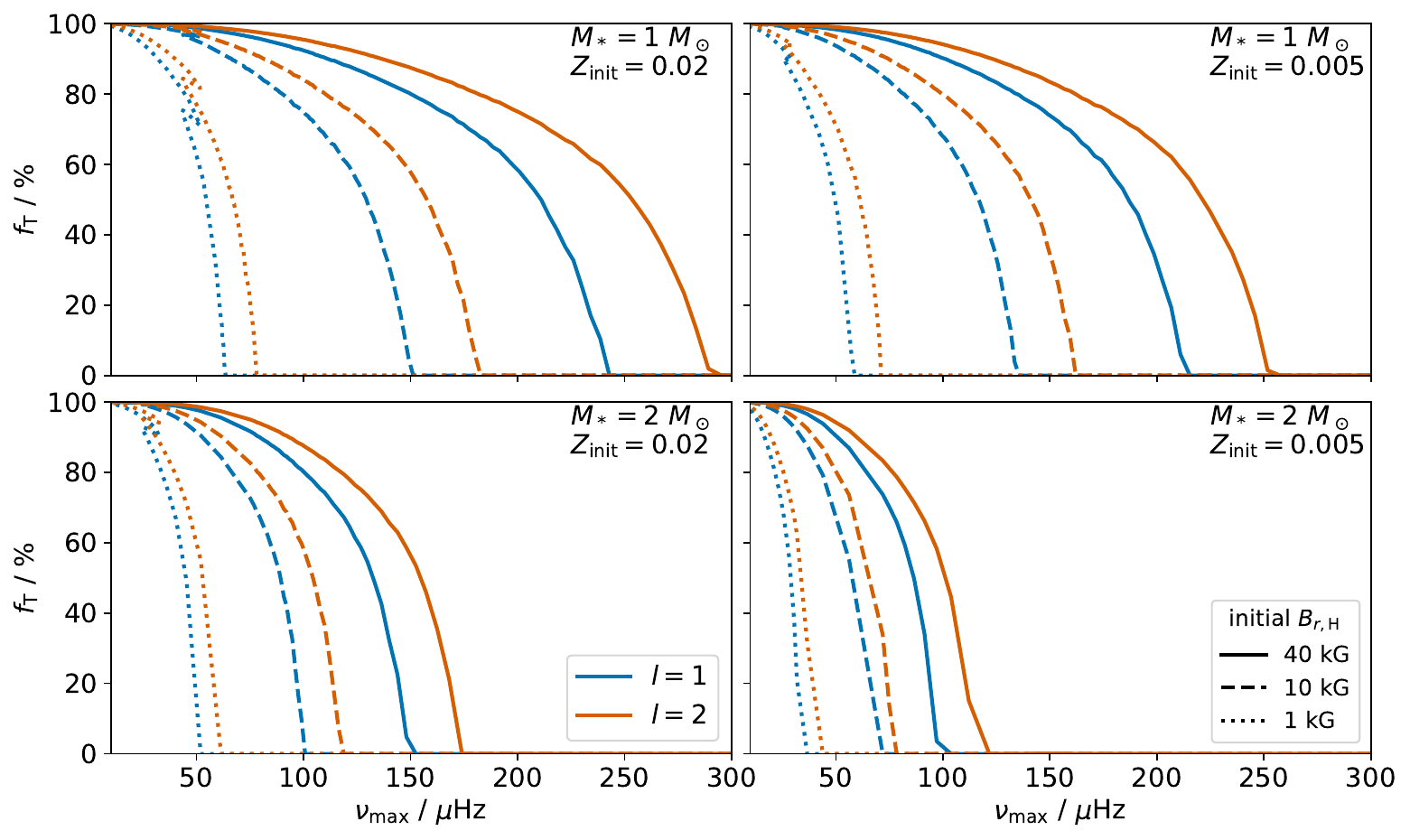}}
    \caption{Same as Fig. \ref{fig: trapped fraction evol center}, now considering the strength of the radial component of the internal magnetic field at the hydrogen-burning shell. Here, we consider the ratio $B_{\rm H}\ /\ B_{\rm crit,H}$ instead of $B_{\rm cen}\ /\ B_{\rm crit,cen}$ when using Eq. \ref{eq: trapped fraction fit}.}
    \label{fig: trapped fraction evol Hshell}
\end{figure*}

In Fig. \ref{fig: trapped fraction evol Hshell}, we show the evolution of the trapped fraction for the four evolutionary tracks presented in Fig. \ref{fig: HRD} considering the ratio of the strength of the internal magnetic field and the critical field strength at the hydrogen-burning shell.

\newpage

\section{MESA inlist} \label{app: MESA inlist}

The MESA inlist used to generate the evolutionary tracks shown in Fig. \ref{fig: HRD} is as follows: \\
\\
\texttt{\&star\_job \\
create\_pre\_main\_sequence\_model=.false.\\
! .true. for PMS model \\
load\_saved\_model = .true.   \\
! .false. for PMS model \\
load\_model\_filename = '2M\_ms.mod'  \\
! load MS or PMS model \\
save\_model\_when\_terminate = .true.\\ 
save\_model\_filename = '2M\_rgb.mod' \\
! chose model name \\
pgstar\_flag=.false. \\
/ \\
\\
\&eos \\
/  \\
\\
\&kap \\
Zbase = 0.02   ! 0.005 for low Z star \\
/  \\
\\
\&controls \\
initial\_mass=2   ! 1 for 1 M\_sun star \\
initial\_z=0.02   ! 0.005 for low Z star \\
log\_directory='LOGS\_RGB'  \\
! chose output directory name\\
 \\
set\_min\_D\_mix = .true. \\
min\_D\_mix = 5d-1 \\
 \\
energy\_eqn\_option='dedt'  \\
use\_gold\_tolerances=.true. \\
 \\
calculate\_Brunt\_N2=.true. \\
\\
terminal\_interval = 10 \\
do\_history\_file = .true. \\
history\_interval = 1 \\
write\_profiles\_flag = .true. \\
profile\_interval = 50 \\
 \\
mesh\_delta\_coeff=0.3d0 \\
varcontrol\_target=1d-5 \\
min\_allowed\_varcontrol\_target=1d-7 \\
max\_years\_for\_timestep=1.d7 \\
/\& \\
}
\\
Note that the above inlist has no stopping condition.
For the pre-MS models, we used the following stopping condition: \\
\\
\texttt{Lnuc\_div\_L\_zams\_limit=0.99d0 \\
stop\_near\_zams=.true. \\
} \\
For the MS models, we used this stopping condition: \\
\\
\texttt{xa\_central\_lower\_limit\_species(1)='h1' \\
xa\_central\_lower\_limit(1)=1d-3 \\
} \\
For the RGB models, we adjusted the \texttt{run\_star\_extras.f90} file such that the computation stops when $\nu_{\rm max}$ is smaller than 10 $\mu$Hz.
We made the MESA inlist and the \texttt{run\_star\_extras.f90} file used in this work publicly available on Zenodo (doi: \href{https://zenodo.org/records/14924586}{10.5281/zenodo.14924586}).

\end{appendix}
\end{document}